\documentclass[
superscriptaddress,prab,10pt,
 amsmath,amssymb,
 aps,
twocolumn,
]{revtex4-2}

\usepackage{graphicx}
\usepackage{dcolumn}
\usepackage{bm}
\usepackage{paralist}
\usepackage{booktabs}
\usepackage{epstopdf}
\usepackage{epsfig}
\usepackage{natbib}
\usepackage{textcase}
\usepackage{bm}
\usepackage{url}
\usepackage{makecell}
\usepackage{tabularx}
\usepackage{multirow}
\usepackage{array,ragged2e}
\newcolumntype{P}[1]{>{\RaggedLeft\arraybackslash}p{#1}}
\newcolumntype{C}[1]{>{\centering\let\newline\\\arraybackslash\hspace{0pt}}m{#1}}
\usepackage{wrapfig}
\usepackage{float}

\usepackage{textcomp}
\usepackage[]{hyperref}
\usepackage{xcolor}
\definecolor{APSBlue}{RGB}{46, 48, 146}
\hypersetup{
    colorlinks,
    linkcolor = APSBlue,
    citecolor = APSBlue,
    urlcolor = APSBlue
}

\begin{document}


\title{Design and early operation of a new-generation internal beam dump for CERN's Super Proton Synchrotron}

\author{A.~Romero Francia}
\author{A.~Perillo Marcone}
\email{antonio.perillo-marcone@cern.ch}
\author{S.~Pianese}

\author{K.~Andersen}
\author{G.~Arnau~Izquierdo}
\author{J.A.~Briz}
\author{D.~Carbajo Perez}
\author{E.~Carlier}
\author{T.~Coiffet}
\author{L.S.~Esposito}
\author{J.L.~Grenard}
\author{D.~Grenier}
\author{J.~Humbert}
\author{K.~Kershaw}
\author{J.~Lendaro}
\author{A.~Ortega~Rolo}
\author{K.~Scibor}
\author{D.~Senajova}
\author{S.~Sgobba}
\author{C.~Sharp}
\author{D.~Steyaert}
\author{F.M.~Velotti}
\author{H.~Vincke}
\author{V.~Vlachoudis}
\author{M.~Calviani}
\email{marco.calviani@cern.ch}
\affiliation{CERN, 1211 Geneva 23, Switzerland}


\begin{abstract}
The Super Proton Synchrotron (SPS) is the last stage in the injector chain for CERN's Large Hadron Collider, and it also provides proton and ion beams for several fixed-target experiments. The SPS has been in operation since 1976, and it has been upgraded over the years. For the SPS to operate safely, its internal beam dump must be able to repeatedly absorb the energy of the circulating beams without sustaining damage that would affect its function. The latest upgrades of the SPS led to the requirement for its beam dump to absorb proton beams with a momentum spectrum from 14 to 450~GeV/$c$ and an average beam power up to $\sim$270~kW. This paper presents the technical details of a new design of SPS beam dump that was installed in one of the long straight sections of the SPS during the 2019--2020 shutdown of CERN's accelerator complex within the framework of the Large Hadron Collider Injectors Upgrade (LIU) Project. This new beam dump has been in operation since May 2021, and it is foreseen that it will operate with a lifetime of 20~years. The key challenges in the design of the beam dump were linked to the high levels of thermal energy to be dissipated---to avoid overheating and damage to the beam dump itself---and high induced levels of radiation, which have implications for personnel access to monitor the beam dump and repair any problems occurring during operation. The design process therefore included extensive thermomechanical finite-element simulations of the beam-dump core and its cooling system's response to normal operation and worst-case scenarios for beam dumping. To ensure high thermal conductivity between the beam-dump core and its water-cooling system, hot isostatic pressing techniques were used in its manufacturing process. A comprehensive set of instrumentation was installed in the beam dump to monitor it during operation and to cross-check the numerical models with operational feedback. The beam dump and its infrastructure design were also optimized to ensure it can be maintained, repaired, or replaced while minimizing the radiation doses received by personnel.

\end{abstract}

\maketitle

\section{Introduction}
The Large Hadron Collider (LHC) Injectors Upgrade (LIU)~\cite{LIU} and High-Luminosity LHC (HL-LHC) era~\cite{HL_LHC} presents unprecedented challenges for beam-brilliance requirements, requiring the upgrade of several devices in the CERN accelerator complex. In this framework, a new-generation internal beam dump, known as TIDVG\#5 (Target Internal Dump Vertical Graphite, version 5), has been designed and manufactured. This device was installed in the long straight section~5 (LSS5) of the CERN Super Proton Synchrotron (SPS) during the Long Shutdown~2 (LS2, 2019--2020). This system is meant to dispose of the SPS's circulating beam whenever necessary, i.e., in case of emergency, during LHC beam setup or filling, during machine development, and for fixed targets (FTs).

\begin{figure}[htb]
    \centering
    \includegraphics[width=0.48\textwidth]{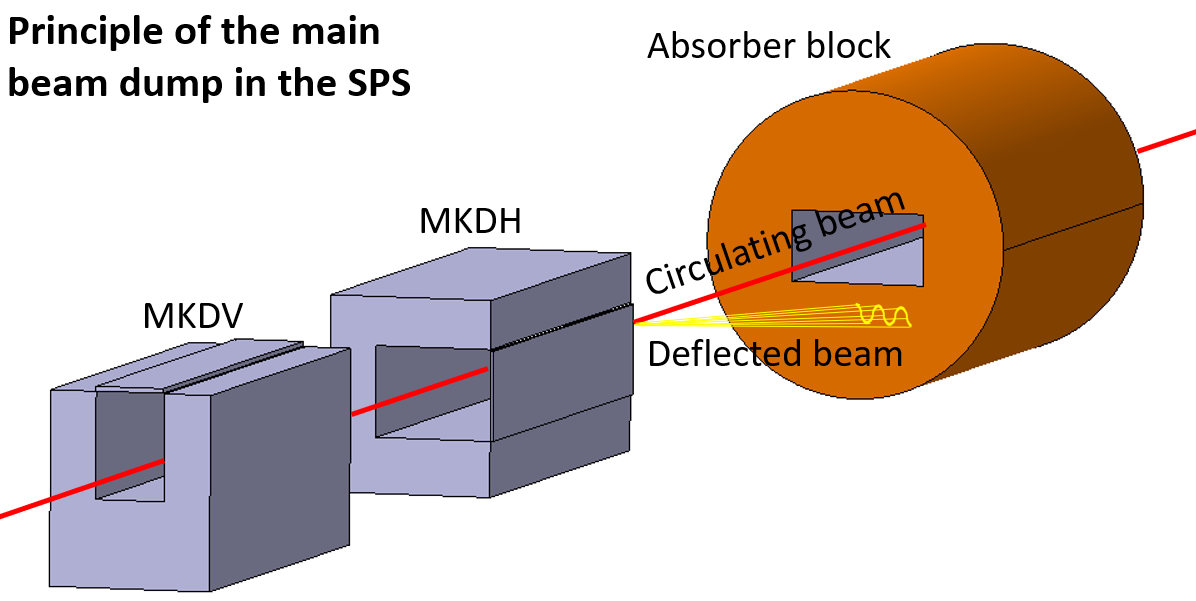}
    \caption{Principle of the main beam dump in the SPS.}
    \label{principle}
\end{figure}

The fundamental concept behind internal beam dumps entails redirecting the beam toward solid blocks, facilitating subsequent energy dissipation through efficient cooling systems. In the specific case of the main internal beam dump of the SPS, whenever high-energy proton beams need to be dumped, they are deflected downward onto the absorbing blocks by a set of three vertical magnetic kickers (MKDVs) and swept horizontally by means of three horizontal magnetic kickers (MKDHs), creating a pattern that dilutes the energy deposited in the dump (see Fig.~\ref{principle}). This beam dilution causes asymmetric deposition of the beam load, which results in one side of the dump to experience higher temperatures than the right side. The thermal power deposited in the blocks is mostly diffused by conduction to CuCr1Zr heat sinks and evacuated through their water circuits.

The core assembly is located inside the accelerator, and it is thus in the same ultra-high vacuum (UHV) environment. Simultaneously, the core is encased within inner and external shields. The primary objective of this shielding arrangement is to confine and prevent the escape of radiation from the core as well as to protect intervening personnel during maintenance periods.

According to the authors’ knowledge, no other similar beam intercepting device can be directly compared with the TIDVG dump block, as it is a high-power, high-energy device that operates directly in the UHV. A dump capable of a similar beam kinetic energy - but external to the machine UHV - operated at the Main Injector at Fermilab~\cite{MI40_505401}. Other high-power external beam dumps are operating at Oak Ridge National Laboratory (ORNL)~\cite{KUMAR2021165380} as well as at the J-PARC Hadron Hall~\cite{Takahashi_2011}. A high-power device that is foreseen to operate in vacuum is the Facility for Rare Isotope Beams (FRIB) ion beam dump~\cite{AVILOV201624}, but it still need to be built and to demonstrate its capability to operate reliably. Alternative design for internal beam dumps can be found also for the tuning dump of the European Spallation Source~\cite{Lee:IPAC2017-THPVA065}, for the RCS at J-PARC~\cite{doi:10.1080/00223131.2022.2038301} as well as the LIPAc facility~\cite{BRANAS2018127}, all of them however operating at a significantly lower beam kinetic energy.

\subsection{History of SPS beam-dump evolution}

From the moment the SPS was commissioned in April 1976, there has been always an internal beam dump. The first version, named the Target Internal Dump Vertical (TIDV), was designed to absorb beams of 10$^{13}$ protons per pulse at 400~GeV/$c$~\cite{tidv}, and it defined the shapes of the next generations of beam dump. The TIDV consisted of a core and a protective iron shielding. The core was composed of two parts: a front part made of aluminium and a downstream part made of copper, in line with the design concept of increasing the density of the core to spread out the energy of the beam along the whole length of the dump. To evacuate the deposited thermal energy, channels were drilled through the core to enable cooling it with water.

Version~II maintained the use of an aluminum core to absorb the beam energy while introducing a surrounding copper core for more efficient heat transfer to the water-cooling channels~\cite{tidv2}. Version~III addressed a significant drawback of prior iterations by removing flexible-bellow connections at the upstream and downstream ends of the beam dump~\cite{tidv3}. These connections caused some issues due to the large vibrations resulting from heavy beam-dumping shocks. Instead, in Version~III, the water-cooling pipes were redesigned with flexible connections that passed through the vacuum tank and the SPS vacuum chamber.

In the early 1990s, when the LHC project was proposed, there were concerns about whether TIDV~III could withstand higher intensities and repetition rates~\cite{tidvg,tidvgg}. To address this, a new generation of internal beam dumps, the TIDVG, was designed. It consisted of two internal dump blocks in long straight section 1 (LSS1) of the SPS: the Target Internal Dump Horizontal (TIDH) for lower energies (14 to 28.9~GeV/$c$ momentum) and the TIDVG for higher energies (102.2 to 450~GeV/$c$ momentum). The TIDVG aimed for a more even heat distribution within its core to reduce beam-induced thermal stresses~\cite{tidvg1}. It was made from 250~cm of graphite followed by 100~cm of aluminum and 30~cm of tungsten. It was surrounded by a copper core with four cooling pipes. The copper core was constructed from two halves that were joined using electron-beam welding to make it leak-tight, and the graphite core was covered with titanium foil to prevent the spread of graphite particles from beam-induced shocks and reduce outgassing.

In 2003, an obstruction was discovered in TIDVG: it was found that the beam had penetrated the titanium foil, causing molten titanium to spread inside the beam aperture. In 2006, TIDVG\#1 was replaced with TIDVG\#2, which retained the same design but eliminated the titanium foil.

\begin{figure}[htb]
    \centering
    \includegraphics[width=0.48\textwidth]{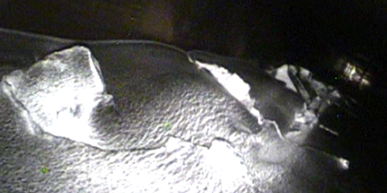}
    \caption{Local melting of aluminium in the absorbing blocks of the TIDVG\#2 beam aperture.}
    \label{melting}
\end{figure}

A subsequent inspection of TIDVG\#2 revealed substantial beam-induced damage, including local melting of the aluminum block~\cite{TIDVG2} (see Fig.~\ref{melting}). This led to its replacement with a modified spare including minor adjustments to the original design, TIDVG\#3~\cite{TIDVG3}. To prevent a recurrence of this issue, operational restrictions were imposed to limit the thermal power deposited in the dump.

In 2014, the TIDVG\#3 was installed with a design similar to its predecessors, using two oxygen-free electronic copper halves bonded together via electron-beam welding. This copper core served both cooling and vacuum-chamber purposes. In 2016, a vacuum leak was identified in the welding between the copper halves, leading to operational limitations. An urgent project was initiated to develop a new dump, resulting in the installation of the TIDVG\#4 between December 2016 and April 2017 during the Extended Year-End Technical Stop~\cite{TIDVG4}. This upgraded device could handle a maximum average power of around 70~kW. In the new design, the copper core was replaced with a CuCr1Zr core enclosed in a stainless-steel (SS) vacuum chamber, eliminating the need for the core itself to serve as a vacuum chamber. The TIDVG\#4 operated reliably until the end of Run~2 in December 2018, as intended. However, considering the anticipated significantly higher intensities and dumping rates for LIU beams---reaching up to around 270~kW---further device upgrades became necessary.

\subsection{Key changes in TIDVG\#5 }
\begin{figure}[htb]
     \centering \includegraphics[width=0.48\textwidth]{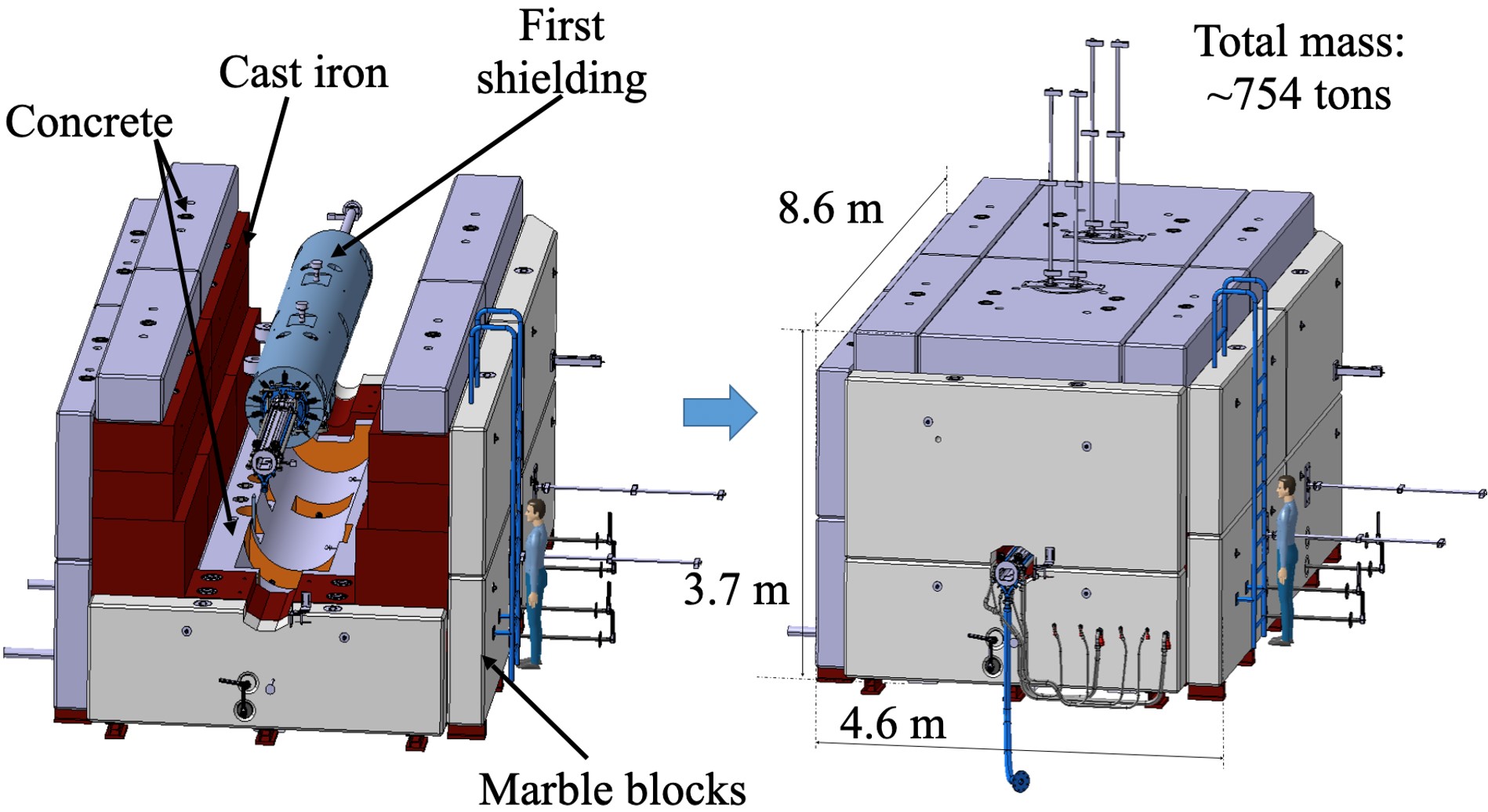}
     \caption{Schematic of TIDVG\#5 assembly, showing the core enclosed in the first shielding and the multi-layer shielding.}
     \label{fig:TIDVG_Assemb}
\end{figure}

A new fifth-generation internal beam dump (TIDVG\#5) was produced and installed during LS2 (2019--2020) (Fig.~\ref{fig:TIDVG_Assemb}). There were several key changes made for this installation. First, the beam dump was relocated to SPS LSS5 to overcome numerous limitations imposed by the LSS1 location. These included the vacuum pressure rise in the SPS injection kicker magnets due to significant outgassing by the TIDVG, reliability issues with various equipment, and the impossibility of completely enclosing the dump in shielding~\cite{LSS5}. A second significant change involved replacing the two previous dumps (TIDH and TIDVG\#4)~\cite{TIDH} with a single device covering the whole range of SPS beam momenta, i.e., from 14 to 450~GeV/$c$.

\begin{figure}[htb]
     \centering \includegraphics[width=0.48\textwidth]{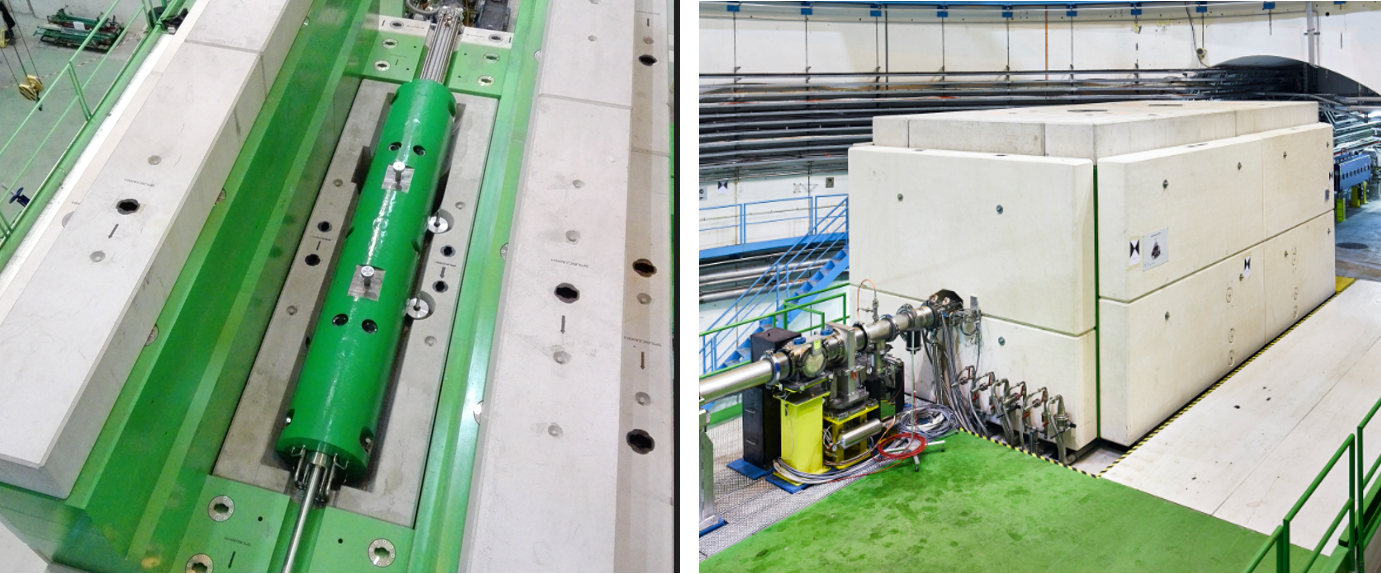}
     \caption{Full TIDVG\#5 assembly, as currently installed in the ECX5 cavern. The left picture shows the dump during the installation process, showing the first shielding. The right picture shows the operational configuration, with the concrete and marble shielding~\cite{Ordan:2742259}.}
     \label{fig:TIDVG_Assemb1}
\end{figure}

TIDVG\#5 is fundamentally different from its predecessors, especially when compared to TIDVG versions 1, 2, and 3. First, it is designed to cope with more than four times the deposited thermal beam power, $\sim$270~kW against the $\sim$60~kW of TIDVG\#4. The total length of graphite was extended to reduce the energy density deposited in the higher-density materials downstream. This---combined with the requirement to maintain a similar equivalent interaction length (an attenuation factor of at least $4.21 \times 10^{-7}$)---increased the core length by 70~cm (making a total of 5.0~m compared to the 4.3-m length of TIDVG\#4). To obtain the highest possible cooling efficiency from the heat sinks, hot isostatic pressing (HIP) was employed to diffusion bond the SS pipes to the CuCr1Zr core. The different beam optics in LSS5 allowed better horizontal centering of the dump with respect to the beam axis. Finally, all previous versions were only equipped with a single layer of iron shielding; in contrast, the external multi-layered shielding of the new dump will contribute to considerably lowering the residual dose rate in the straight-section area where the dump is installed (see Figs.~\ref{fig:TIDVG_Assemb} and \ref{fig:TIDVG_Assemb1}).

\subsection{Scope of this paper}
This paper presents the detailed design, material selection, and manufacturing techniques associated with the main sub-assemblies of this new dump: the core absorbing blocks, the HIP cooling plates~\cite{HIP_Cu}, the air-cooled vacuum chamber, and the external multi-layered shielding, which is made from reinforced concrete, cast iron, and white marble, making a total weight of almost 754~t. The entire assembly is 8.6~m long, 3.7~m high, and 4.6~m wide.

Using the framework of this design, detailed thermomechanical finite-element method (FEM) simulations were carried out to examine the dump core's performance, taking into account operational and worst-case scenarios for beam operation. Finally, details of the instrumentation that was installed to monitor the behavior of the dump in operation and benchmark the numerical models (temperature sensors, flow meters, and linear variable differential transformers (LVDTs)) are presented and described in detail here.

\section{TIDVG\#5 absorber design}
\label{sec:Core}
\begin{figure}[ht]
    \centering
    \includegraphics[width=0.5\textwidth]{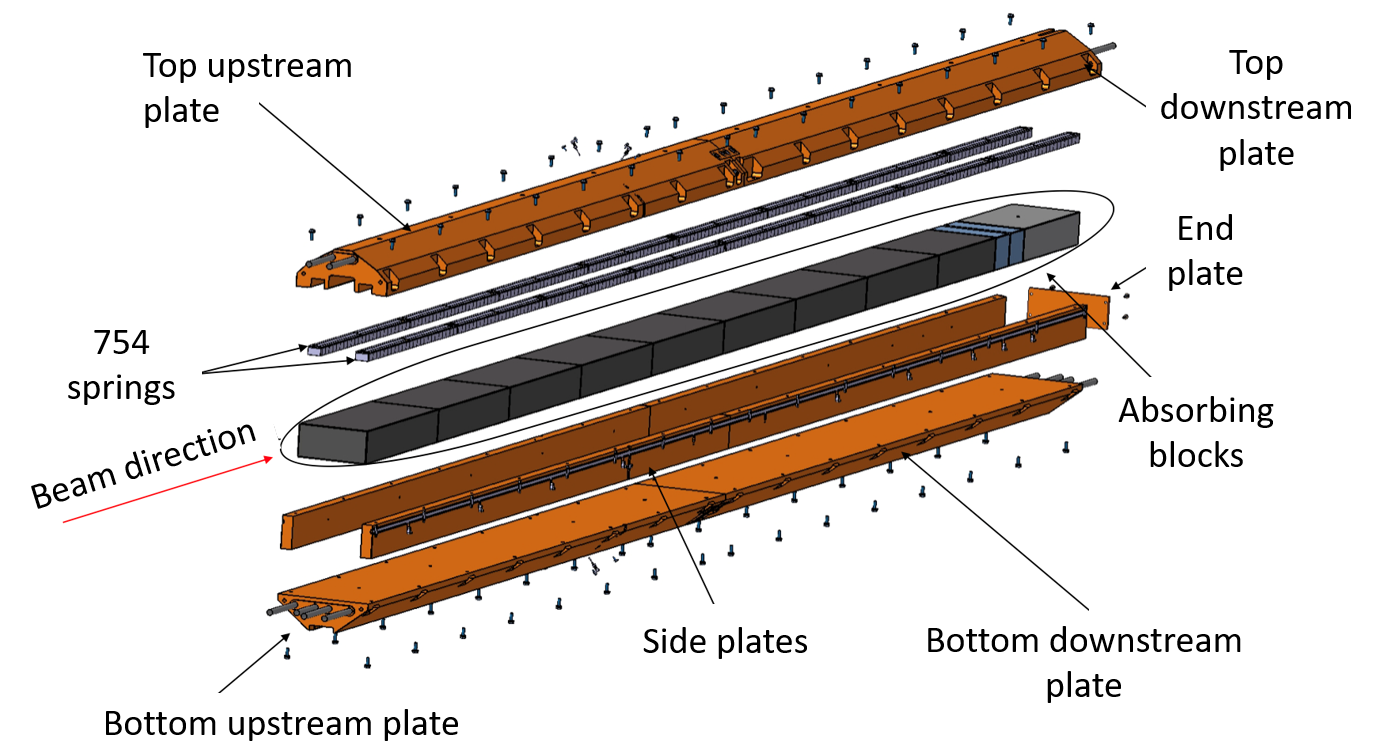}
    \caption{Exploded schematic showing plate assembly. In order following the beam direction: graphite (dark grey), TZM (light blue), and tungsten (light grey); the CuCr1Zr cooling plates and compression springs for the cooling of the blocks are also shown.}
    \label{figCore1}
\end{figure}

\begin{figure}[ht]
    \centering
    \includegraphics[width=0.5\textwidth]{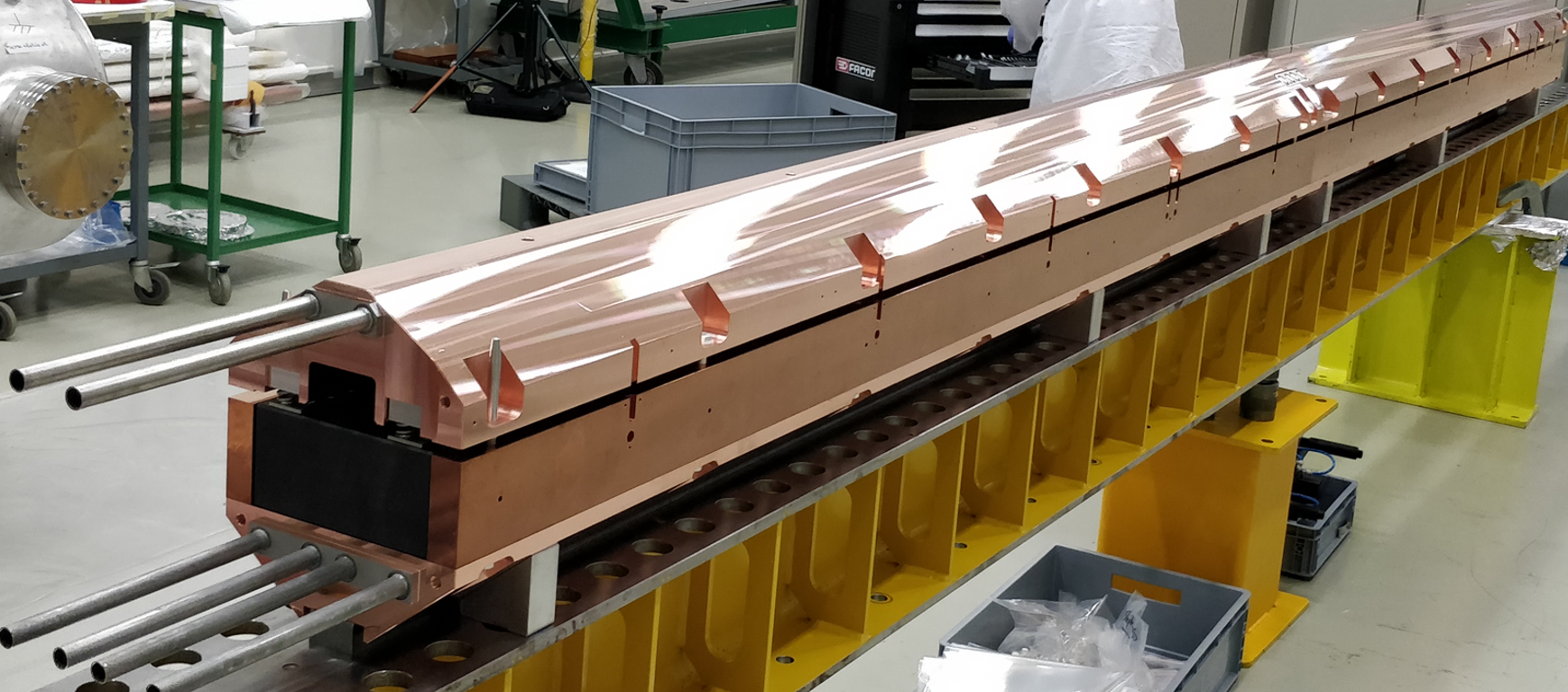}
    \caption{Photograph of the assembly of the absorbing blocks in the CuCr1Zr core in the clean room.}
    \label{realcore}
\end{figure}

The active part of the TIDVG\#5 consists of an array of absorbing blocks made of 4.4~m of isostatic graphite, 0.2~m of titanium--zirconium--molybdenum alloy (TZM) and $\sim$0.4~m of pure tungsten. These are enclosed by an assembly of water-cooled (top/bottom) and non-cooled (side) CuCr1Zr plates, as shown in Figs.~\ref{figCore1} and \ref{realcore}. The absorbing blocks are pressed against the bottom cooling plates to evacuate the energy deposited by the particle beam.

\begin{figure}[ht]
    \centering
    \includegraphics[width=0.5\textwidth]{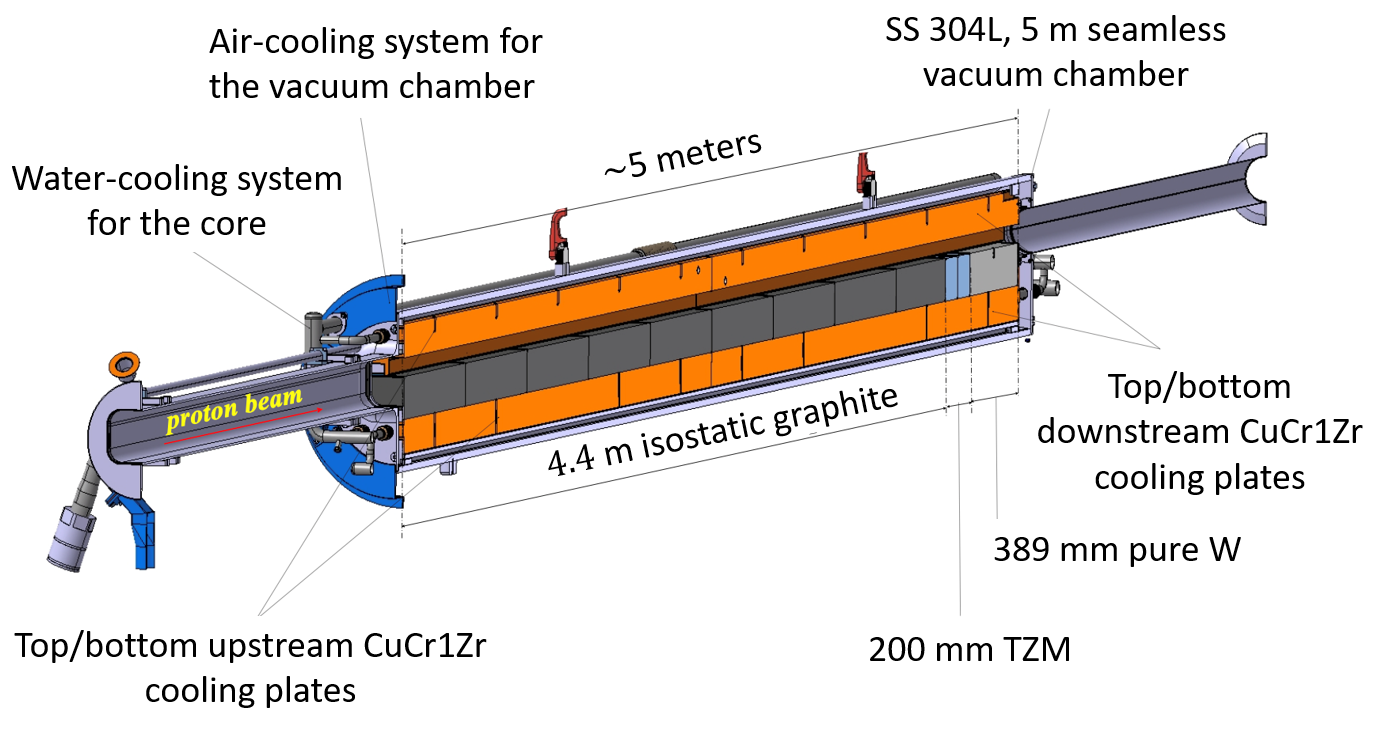}
    \caption{Cross section of the TIDVG\#5 SPS internal dump core; the cast-iron first shielding is hidden for clarity.}
    \label{fig:CrossSectionTIDVG}
\end{figure}

A 5-m-long, seamless, multi-directionally forged 304L SS vacuum chamber encloses the dump core (Fig.~\ref{fig:CrossSectionTIDVG}).

Cast-iron shielding, comprising two cylindrical blocks (top and bottom shielding) weighing 8~t each, is assembled around the vacuum chamber containing the dump core. Additionally, the top shielding can be easily manipulated by means of two lifting points.

The vacuum chamber is cooled by a flow of air generated by a dedicated ventilation system. Air at room temperature is extracted from the cavern in which the TIDVG\#5 is installed and continuously channeled into a 10-mm gap between the chamber and the first shielding. To further contain the radiation from the dump core, a massive, multi-layered external shielding system completes the dump-block assembly.

\subsection{Absorbing blocks}
The combination of materials and lengths of the absorbing blocks is of paramount importance for minimizing the density of the energy deposited by the beam and therefore keeping the stresses associated with the resulting thermal gradients within acceptable ranges. Table~\ref{tab1} lists the layout, densities, and lengths of the TIDVG\#5 core materials.

\begin{table}[h!t]
    \centering
    \caption{TIDVG\#5 core materials.}
    \begin{tabular}{C{2.5cm} C{1.5cm} C{2cm} C{2cm}}
        \toprule \toprule
        Material & Density (g/cm$^{3}$) & Number of blocks & Block length (cm) \\
        \midrule
        Graphite R7550 & \multicolumn{1}{c}{1.8} & \multicolumn{1}{c}{9} & \multicolumn{1}{c}{440} \\
        TZM & \multicolumn{1}{c}{10.1} & \multicolumn{1}{c}{2} & \multicolumn{1}{c}{20} \\
        Tungsten & \multicolumn{1}{c}{18.8} & \multicolumn{1}{c}{1} & \multicolumn{1}{c}{39} \\
        CuCr1Zr & \multicolumn{1}{c}{8.9} & \multicolumn{1}{c}{1} & \multicolumn{1}{c}{1} \\
        \bottomrule \bottomrule
    \end{tabular}
    \label{tab1}
\end{table}

The 4.4~m of isostatic graphite is divided into eight blocks of $200 \times 96 \times 500$~mm$^3$ (width $\times$ height $\times$ length) and one block of $200 \times 96 \times 400$~mm$^3$. The two TZM blocks are $200 \times 95 \times 100$~mm$^3$, and the tungsten block is $200 \times 95 \times 389$~mm$^3$. The array ends with a 1-cm-thick CuCr1Zr plate. With this configuration, the new dump is improved with respect to TIDVG\#4 in terms of the survival probability factor (i.e., the ratio between the primary uncollided particles escaping and those impinging on the dump), which is $1.7 \times 10^{-7}$ for TIDVG\#5 and $4.2 \times 10^{-7}$ for TIDVG\#4.

The absorbing blocks are arranged so that their densities increase as the beam passes through the device. The firsts blocks, made of graphite, are meant to dilute the beam and reduce the energy density deposited in the higher-density materials. In contrast, the latter blocks are designed to protect the downstream hardware from the particle shower escaping from the graphite blocks. This material is a fine-grain isostatic graphite (SGL Carbon R7550~\cite{sgl}). The TZM blocks (produced by AT\&M~\cite{atm}) are expected to be the most thermomechanically loaded components. To achieve high mechanical strength, they were 2D forged to obtain a higher ductility in the transverse direction of the block and an overall isotropic microstructure in the final part. For the tungsten block, forging it would have offered a result with superior mechanical properties, but this was not needed due to the small amount of beam load to be deposited on this block. The tungsten block in its pure form, sourced from AT\&M, underwent a cost-effective manufacturing process involving sintered pre-shapes to minimize production expense. Subsequently, it underwent further densification through HIP to achieve an optimal density of $\sim$97\% from its nominal value.

The tensile strengths of TZM and tungsten can be found in Table~\ref{tab:Strength}. The manufacturer conducted tests on both materials, and the former underwent additional testing at CERN. For this particular material, during the testing campaign, samples were extracted from a spare block in three different directions. The absorbing blocks operate in a UHV environment and must comply with severe cleanliness standards. In contrast, the graphite blocks were dry-machined (no lubricant was used) and thermo-chemically purified under an argon atmosphere to keep the ash content below 5~ppm. Before installation, the blocks were vacuum-fired, i.e., heated to 950$^{\circ}$C in a vacuum furnace with a dwell time of 16~h; the TZM and tungsten blocks were treated at 1000$^{\circ}$C for 6~h.

\begin{figure}[htb]
    \centering
    \includegraphics[width=0.48\textwidth]{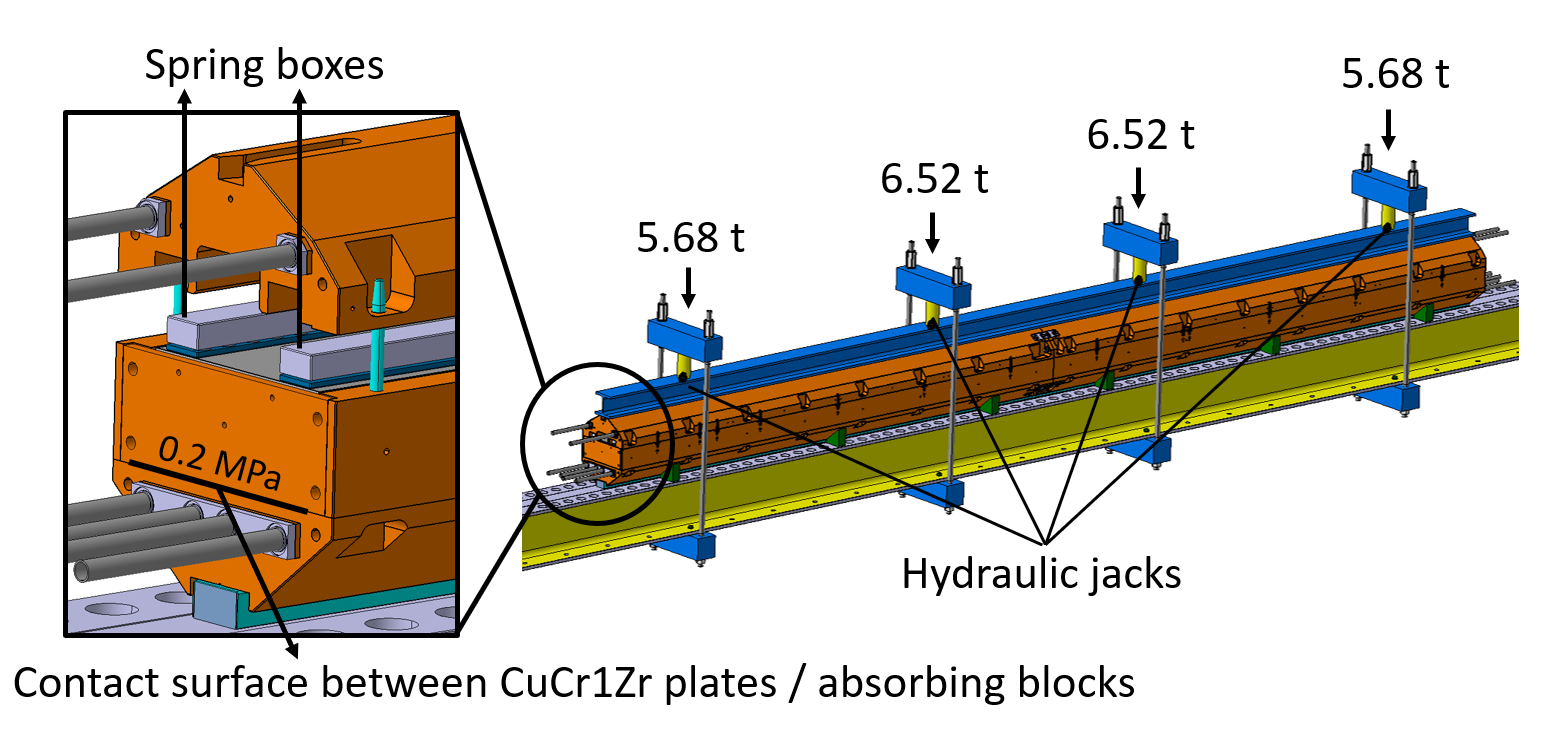}
    \caption{On the left-hand side, an exploded view of the CuCr1Zr core illustrates how the spring boxes are installed. The right-hand illustration shows how the absorbing blocks are pushed against the CuCr1Zr plates.}
    \label{fig:CP1}
\end{figure}

To reach an efficient thermal contact conductance (TCC) at the interface between the absorbing blocks and the CuCr1Zr plates, two sets of 377 springs on each side of the beam aperture are installed in spring boxes. By means of two hydraulic jacks at the extremities (exerting a force of 5.68~t each) and two hydraulic jacks in the middle (6.52~t each), the absorbing blocks are pushed against the CuCr1Zr plates with a force of 24.4~t, as shown in Fig.~\ref{fig:CP1}. The contact pressure is ensured by determining the stiffness of the springs and their deformation during compression.

The estimated contact pressure generated at the interface between the absorbing blocks and the bottom upstream/downstream plates amounts to 0.2~MPa. The contact surfaces of the blocks are precisely machined with a surface roughness of $R_{a}<1.6$~\textmu m and a flatness of 0.04~mm.

\begin{table}[htb]
\caption{Measured ultimate tensile strengths of the TZM and tungsten blocks at room temperature. Quality certification provided by AT\&M, measured on sample coupons. For TZM, B indicates the forging in the beam direction, A indicates the horizontal direction and C the vertical one.}
\begin{tabular}{c m{1.5cm} m{1.5cm} m{1.5cm}}
\toprule \toprule
\multicolumn{1}{l}{} & \multicolumn{3}{c}{Ultimate tensile strength (MPa)} \\ \hline
\multirow{2}{*}{TZM}& \hspace{5.5mm} A & \hspace{6mm}B & \hspace{6mm}C \\
\cline{2-4}  & \hspace{5.4mm}665 & \hspace{5mm}683  & \hspace{5mm}721 \\
\hline
W    & \multicolumn{3}{c}{550}                                 \\
\bottomrule \bottomrule
\end{tabular}
\label{tab:Strength}
\end{table}

\subsection{CuCr1Zr cooling plates}
\subsubsection{Design and materials}
In the worst-case operational scenario, the total thermal power carried by the particle beam can be as high as $\sim$270~kW, and approximately 70\% of this has to be dissipated by the cooling plates of the core. This fraction accounts for the energy directly deposited in the plates by the particle beam shower as well as that evacuated by conduction from the absorbing blocks to the CuCr1Zr heat sink. Hence, the heat-evacuation efficiency of the cooling plates is crucial. The cooling plates are made from CuCr1Zr bonded with 316L SS tubes by means of HIP; these were specially developed for this device~\cite{HIP_Cu}.

\begin{figure}[htb]
\centering
    \includegraphics[width=0.48\textwidth]{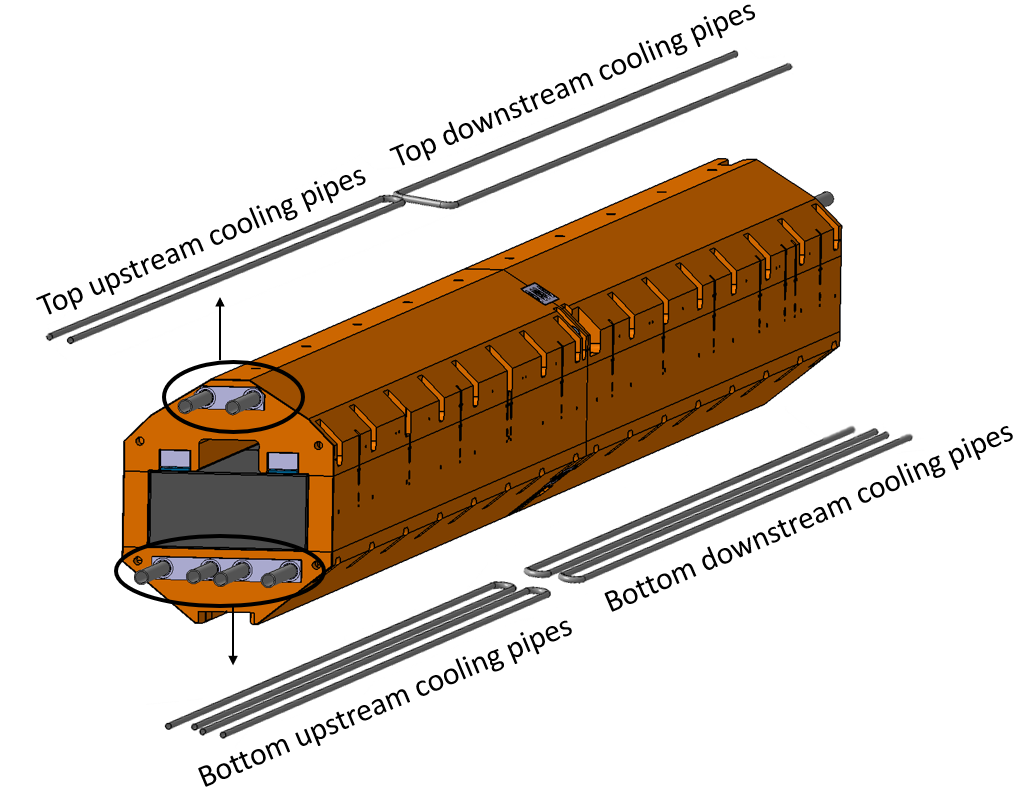}
    \caption{Cooling circuits within the TIDVG\#5 core.}
    \label{fig:Circuits}
\end{figure}

Due to size constraints linked to the HIP process, the cooling plates were divided into four parts: two bottom upstream/downstream plates ($265 \times 93.5 \times 2500$~mm$^3$) and two top upstream/downstream plates ($265 \times 118 \times 2500$~mm$^3$). The top and bottom plates are bolted to the four side plates ($95 \times 31.5 \times 2500$~mm$^3$), keeping them together (Fig.~\ref{fig:Circuits}).

The CuCr1Zr (provided by Zollern GmbH (Germany)) is heat treated and multi-directionally forged according to the EN~12420 standard. The heat treatment involves solution annealing at 950--1000$^{\circ}$C for 30~min followed by water quenching and precipitation hardening through aging, maintaining at 480--500$^{\circ}$C for 2~h, and then slow cooling in air (this treatment will hereafter be referred to as SA+WQ+A). After the HIP cycle, due to the high temperature to which the material has been exposed, the same thermal treatment is repeated to recover the mechanical and thermal properties. In these conditions, CuCr1Zr features high electrical and thermal conductivity as well as high hardness, tensile properties, ductility, and machinability. Table~\ref{tab:Tab2} summarizes the main thermophysical and mechanical properties of CuCr1Zr after SA+WQ+A treatment. These values were measured \textit{in situ} at CERN for its predecessor, TIDVG\#4, and they were used for the thermomechanical simulations in the new version.

\begin{table}[htb]
    \centering
    \caption{Measured thermo-physical and mechanical properties of solution-annealed, water-quenched, and precipitation-hardened through aging of CuCr1Zr at room temperature.}
    \begin{tabular}{m{5cm} m{3cm}}
    \hline \hline
    \multicolumn{2}{c}{Properties of CuCr1Zr at room temperature} \\
         \hline
         {Density} & 8.89~g/cm$^{3}$ \\
         {Specific heat capacity} & 0.37~J/(g\,K)\\
         {Coefficient of thermal expansion} &  $16.3 \times 10^{-6}$~K$^{-1}$ \\
         {Thermal conductivity} & 324~W/(m\,K) \\
         {Young's modulus} & 135~GPa\\
         {Yield strength} & 280~MPa\\
         {Ultimate strength} & 400~MPa \\
         \hline \hline
    \end{tabular}
    \label{tab:Tab2}
\end{table}

The 316L SS cooling tubes were bent at room temperature and then annealed. They have an inner diameter of 15~mm and a wall thickness of 1.5~mm. The annealing cycle for 316L SS includes a heating phase at 200--300$^{\circ}$C/h up to 950$^{\circ}$C, followed by maintenance at this temperature for at least 2~h and then natural cooling to ambient temperature.

The CuCr1Zr is cooled during operation by a total flow of 15~m$^3$/h of demineralized water distributed in six parallel circuits. By design, the temperature rise of the water should not exceed $\sim$20$^{\circ}$C. Since the inlet and outlet of the cooling pipes are relatively close to each other, a larger thermal gradient could induce high thermomechanical stresses. However, the maximum outlet water temperature is set to be 45$^{\circ}$C; higher outlet temperatures would decrease the overall cooling efficiency.

Figure~\ref{fig:Circuits} shows the bottom and top SS pipes. The cooling circuits in the top CuCr1Zr plates will handle a flow rate of 2.5~m$^3$/h each, dissipating 30~kW in the upstream and 14~kW in the downstream. For the bottom plates, the cooling circuits will manage a flow rate of 5~m$^3$/h both upstream and downstream, dissipating 96 and 47~kW, respectively.

\subsubsection{Manufacturing process}
\begin{figure}[htb]
\centering
    \includegraphics[width=0.48\textwidth]{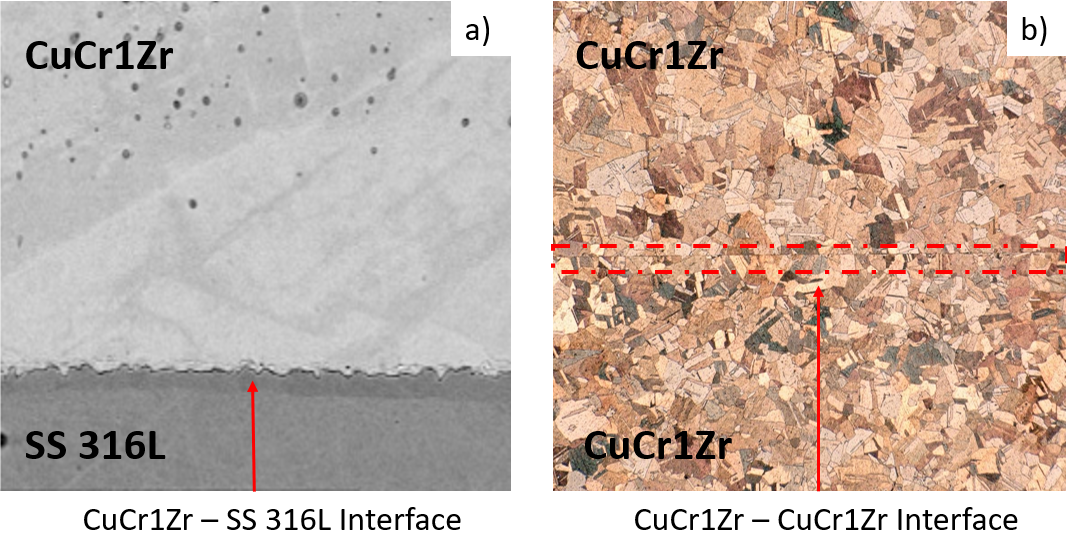}
    \caption{a) SEM image 500x of a CuCr1Zr-SS 316L interface, b) Optical microscopy image of a sample containing the HIP diffusion bonded CuCr1Zr-CuCr1Zr interface~\cite{HIP_Cu}.}
    \label{fig:Interfaces}
\end{figure}

\begin{figure}
    \centering
    \includegraphics[width= 0.48\textwidth]{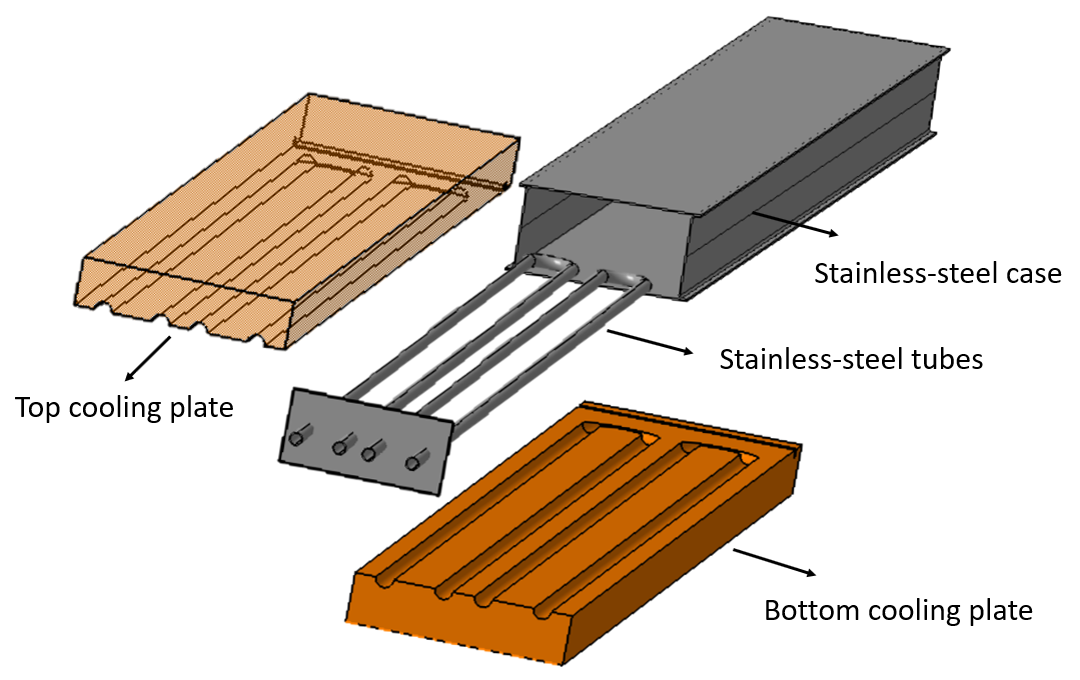}
    \caption{Principle for producing the CuCr1Zr cooling plates. The top and bottom cooling plates are bonded to the SS tubes during the HIP cycle.}
    \label{fig:Before_HIP}
\end{figure}

Diffusion bonding by HIP was employed to produce the CuCr1Zr cooling plates, as illustrated in Fig.~\ref{fig:Before_HIP}. By combining temperatures up to $\sim$950$^{\circ}$C and argon pressure up to $\sim$105~MPa, the interfaces between the 316L SS cooling tubes and the CuCr1Zr, as well as between the top and bottom plates, are perfectly bonded through inter-diffusion of materials, thus virtually eliminating the thermal contact resistance, Fig.~\ref{fig:Interfaces}. The thermal resistance at all interfaces are virtually removed. Likewise, the bonding strength is at least as high as that of the weakest material of the pair~\cite{HIP_Cu}. Experience shows that a gap of around 0.1~mm is best suited to achieving a successful diffusion-bonding process between the materials (Fig.~\ref{fig:adapted_groove}).

The manufacturing process can be sub-divided into three main phases, a) machining and assembly before HIP, b) the HIP cycle and c) thermal treatments and final machining after HIP.

Before the HIP treatment stage, each cooling plate is made from three parts: the cold-bent and annealed 316L SS cooling tubes and the CuCr1Zr top/bottom halves. During the HIP cycle, the gas pressure is exerted inside the tubes and outside the capsule containing the assembly. For the treatment to be successful and to achieve diffusion bonding, these need to be precisely assembled and kept vacuum tight throughout the duration of the cycle.

\begin{figure}
    \centering
    \includegraphics[width= 0.4\textwidth]{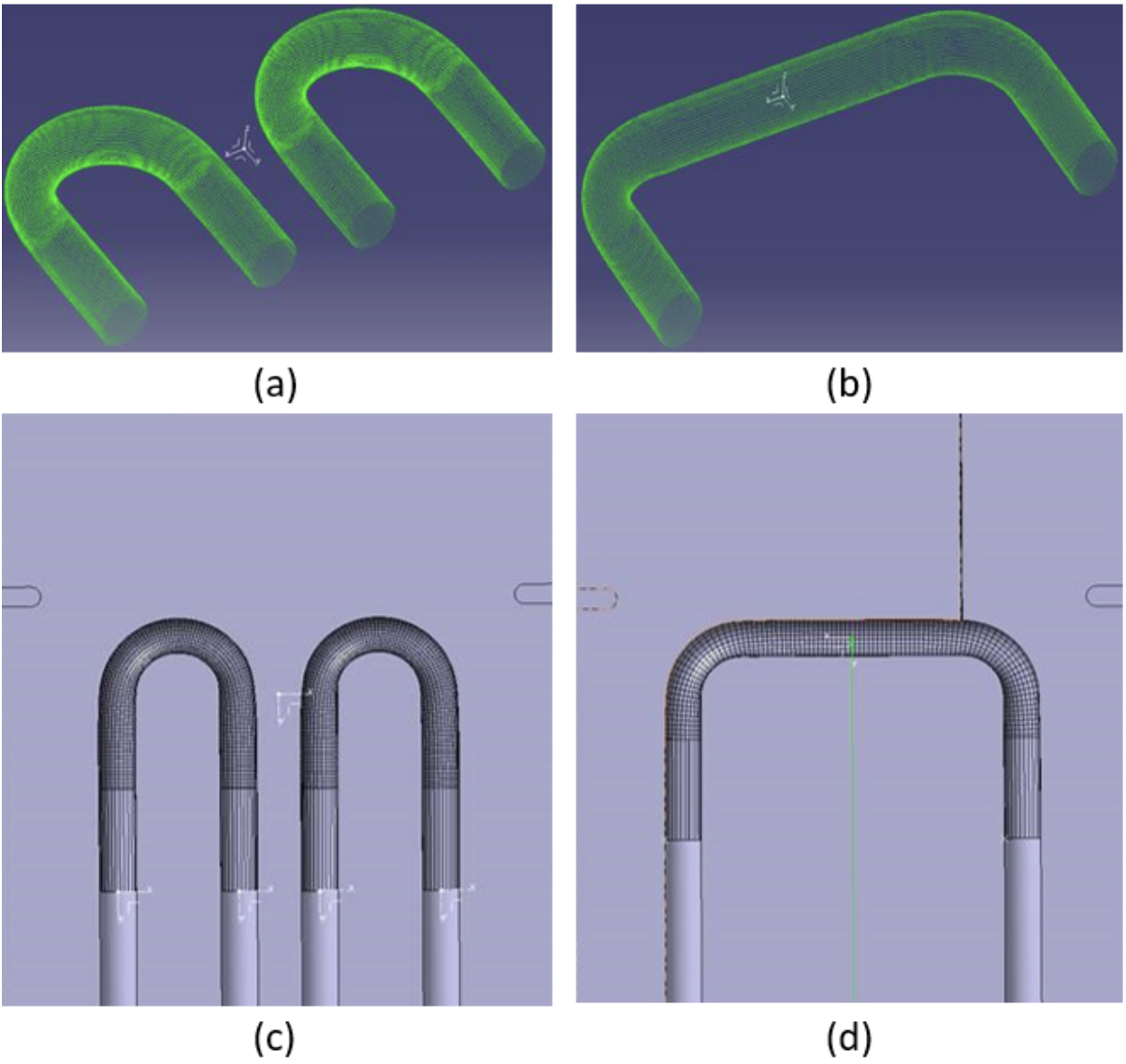}
    \caption{3D point clouds generated for the (a)~bottom and (b)~top downstream plates, with panels (c) and (d) showing the corresponding CAD models.}
    \label{fig:3D_Scan}
\end{figure}

\begin{figure}
    \centering
    \includegraphics[width= 0.4\textwidth]{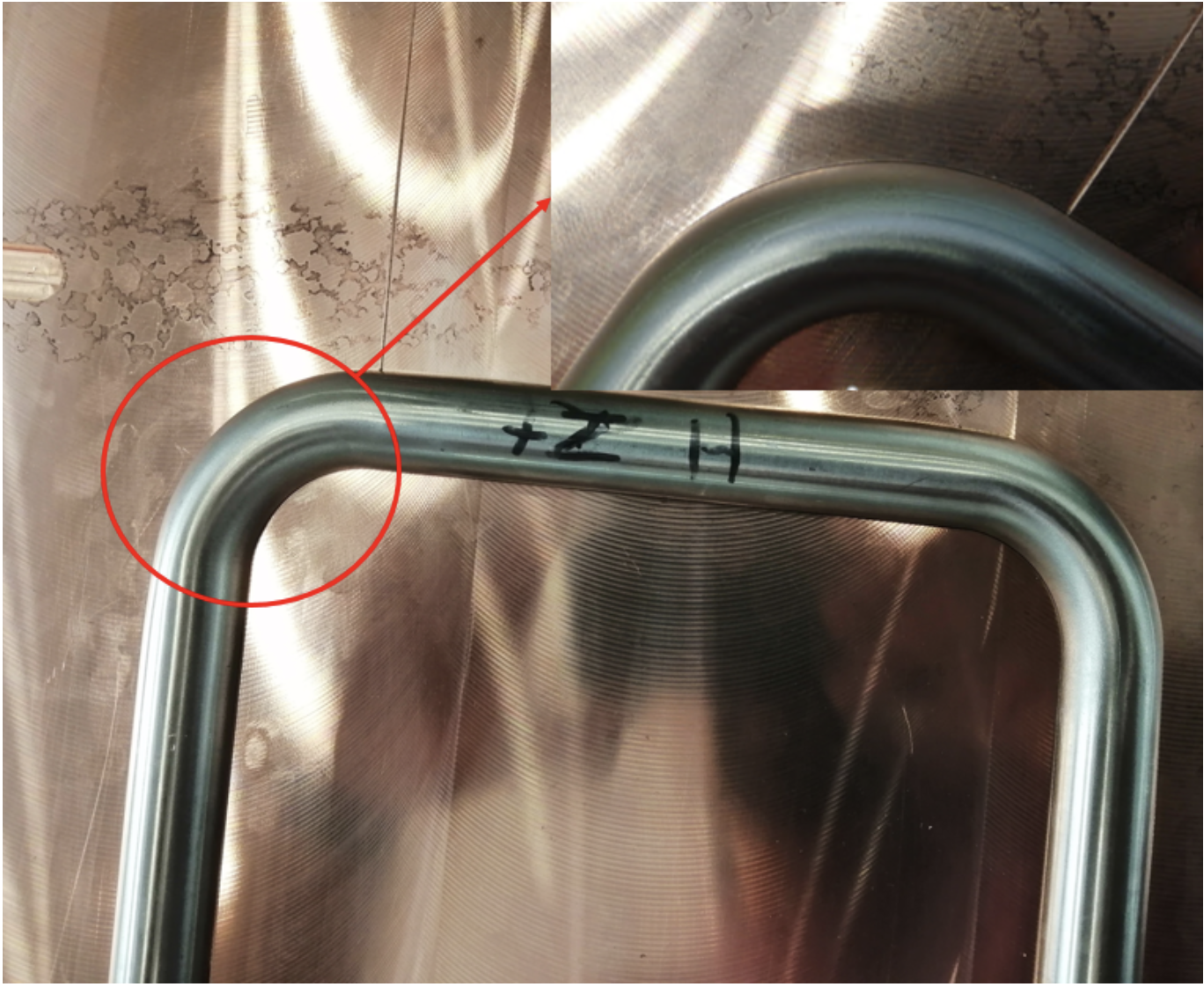}
    \caption{Groove in CuCr1Zr plate machined with a 0.1-mm gap surrounding the tube's bend region.}
    \label{fig:adapted_groove}
\end{figure}

\begin{figure}
    \centering
    \includegraphics[width= 0.48\textwidth]{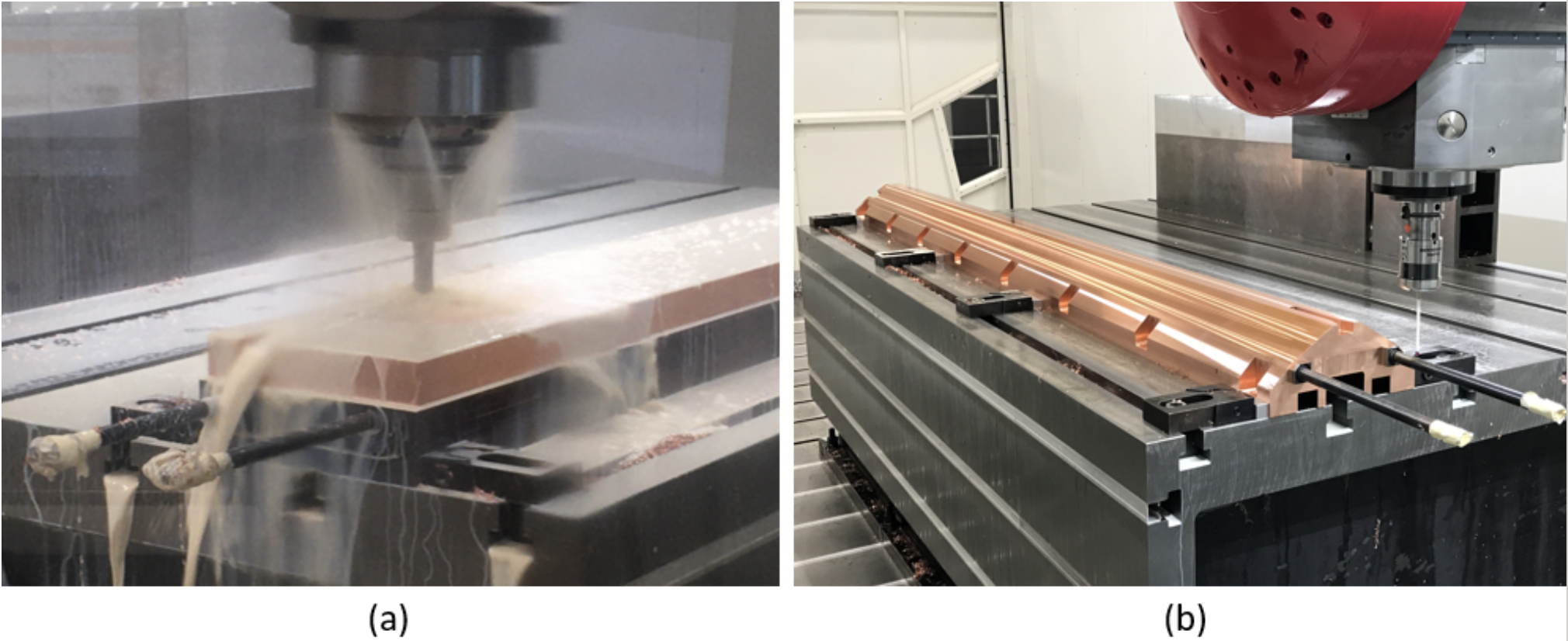}
    \caption{Top downstream CuCr1Zr plate: (a)~surplus material being removed, (b)~final plate.}
    \label{fig:Final_Machining}
\end{figure}

The cold-bending process of the tubes causes a profile distortion with respect to the ``theoretical'' outer surface in the bent regions, with radial offsets of the order of 1~mm. If the gaps between the tubes and the copper grooves are too large, the gas pressure will produce an excessive expansion of the tube during the HIP process, resulting in a high risk of generating local cracks. If this happens, the diffusion bonding will fail for the entire part, as the components will no longer be pressed against each other. To avoid this, the actual shape of each tube was 3D scanned (by means of a coordinate measuring machine (CMM); see Fig.~\ref{fig:3D_Scan}) around the bends to reproduce the same profile (plus a 0.1-mm offset; see Fig.~\ref{fig:adapted_groove}) for the grooves, which were machined in the CuCr1Zr using computer numerical control (CNC); see Fig.~\ref{fig:Final_Machining}.

The HIP pressure/temperature cycle that promotes the diffusion bonding can be divided in four phases. During the first phase, the temperature remains unchanged while the pressure is raised to $\sim$22.5~MPa. In the second (heating) phase, the temperature and pressure are both linearly increased to $\sim$950$^{\circ}$C and $\sim$105~MPa, respectively. In the third or ``dwell'' phase, both parameters are kept constant over 180~min. The last phase corresponds to the cooling period, during which the temperature decreases at a rate of $\sim$5$^{\circ}$C/min (see Fig.~5 of Ref.~\cite{HIP_Cu}). 

During the HIP cycle, because of the exposure to high temperatures, the CuCr1Zr experiences a significant loss of thermophysical and mechanical properties. To restore these properties, the part is again submitted to SA+WQ+A thermal treatment.

For the thermal treatments to be more effective, the SS cladding, which is diffusion-bonded to the underlying CuCr1Z after HIP, was removed by machining. The first thermal treatment, the water-quenching process, produces a bending of the plate of the order of 1~cm. The part was straightened before final machining by means of a hydraulic press. To take into account the machining precision, the deformations induced by thermal treatments, and possible stress relaxations upon final machining, each CuCr1Zr plate had $\sim$10~mm of surplus material on each face to achieve the required final tolerances (Fig.~\ref{fig:Final_Machining}).

\subsection{Vacuum chamber}
The vacuum chamber is another critical component of the SPS internal beam dump. Because it encloses active parts such as the absorbing blocks and the CuCr1Zr water-cooled plates, it must remain leak tight and maintain a UHV environment for the entire lifetime of the dump, including during the most demanding mechanical loads. The chamber is a 5-m-long, seamless 304L SS tube with 15-mm-thick walls.

\begin{figure}
    \centering
    \includegraphics[width= 0.48\textwidth]{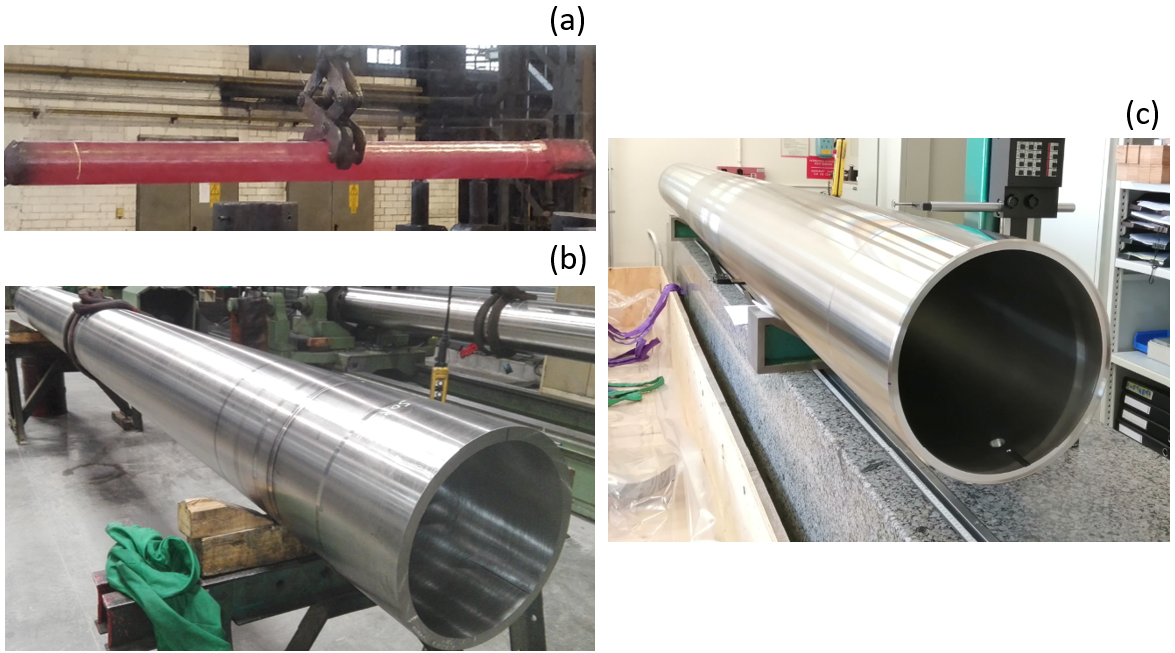}
    \caption{(a)~Cylindrical 304L SS bar being multi-directionally forged, (b)~pre-machined tube, (c)~final machined tube being controlled in the metrology laboratory.}
    \label{fig:Vacuum_Chamber}
\end{figure}

A dedicated manufacturing process was designed and implemented to produce these challenging components. This involved three separate steps (Fig.~\ref{fig:Vacuum_Chamber}): (i)~multi-directional forging of a cylindrical 304L SS bar, (ii)~core boring to obtain a seamless pre-machined tube, and (iii)~final machining to the required dimensions: 5151-mm length with a 0.5-mm straightness tolerance, and inner and outer diameters of 329 and 259~mm respectively; all linear measurements had a tolerance of ±0.5~mm.

The SS employed for this application, 304L, is completely austenitic and homogeneous, without segregations or intermediate phases. To avoid inter-granular leaks, the chamber material has a fine grain structure, with a grain-size number at least 2 according to ASTM~E112. Finally, the material soundness was inspected following the EN~10228-4 standard, ensuring compliance with EN~4050-4 Class~3. To limit the radiological hazard around the dump linked to the activation of the chamber during operation, the maximum admissible cobalt content was set at 0.1\%. Tables ~\ref{Comp} and ~\ref{tab:Table_5} show the results of chemical-composition analysis of the product and its main tensile properties at room temperature, respectively.

\begin{table}[htb]
    \centering
    \caption{Tensile properties at room temperature for the 304L SS of the vacuum chamber. Quality certification provided by Schmiedewerke Gr\"oditz GmbH, GMH Gruppe, measured on sample coupons.}
    \begin{tabular}{C{2cm} c c c c}
    \hline \hline
       Test T ($^{\circ}$C) & $R_{p0,2}$\,(N/mm$^{2}$) & $R_m$ (N/mm$^{2}$) & A5 (\%) & Z (\%)\\
    \hline
    23 & 208 & 487 & 64.0 & 82 \\
    \hline \hline
    \end{tabular}
    \label{tab:Table_5}
\end{table}

\begin{table*}[]
    \centering
    \caption{Chemical-composition analysis of the 304L SS (1.4306) of the vacuum chamber. Quality certification provided by Schmiedewerke Gr\"oditz GmbH, GMH Gruppe, measured on sample coupons.}
    \begin{tabular}{C{1.5cm} C{1cm} C{1cm} C{1cm} C{1cm} C{1cm} C{1cm} C{1cm} C{1cm} C{1cm} C{1cm} C{1cm} C{1cm} C{1cm} C{1cm}}
    \hline \hline 
         & C & Si & Mn & P & S & Cr & Ni & Mo & V & Cu & Al & Co & Bo & N \\
    \hline
    Analysis Product & 0.016 & 0.28 & 1.81 & 0.022 & 0.004 & 17.44 & 11.23 & 0.10 & 0.048 & 0.10 & 0.020 & 0.02 & 0.0006 & 0.014 \\ 
    \hline \hline
    \end{tabular}
    \label{Comp}
\end{table*}

\subsubsection{Air-cooling system}
\begin{figure}
    \centering
    \includegraphics[width= 0.4\textwidth]{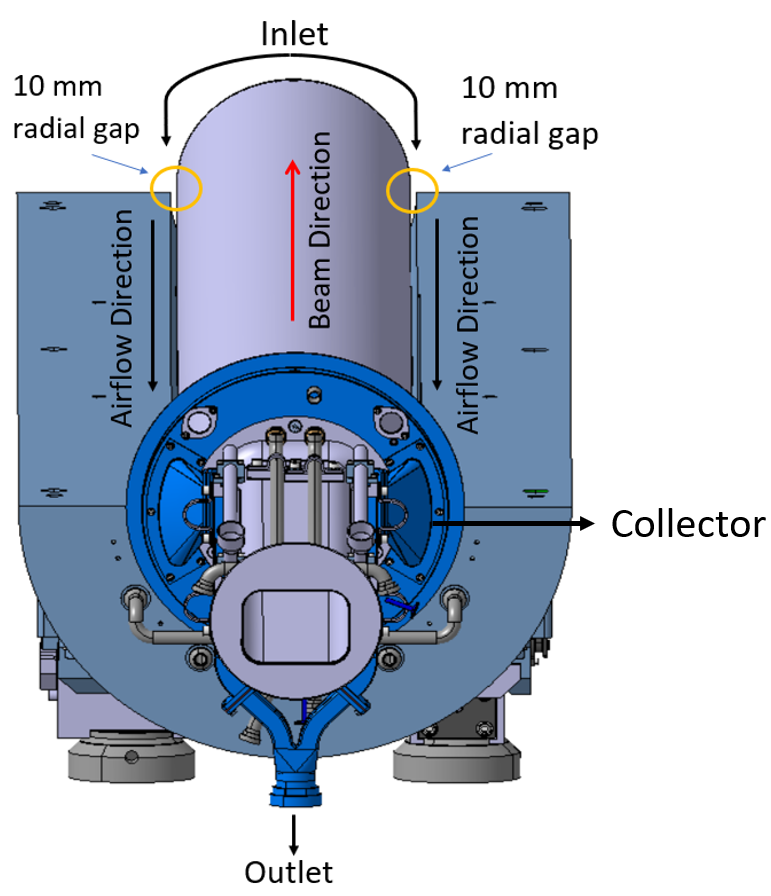}
    \caption{Schematic representation of the air-cooling system surrounding the vacuum chamber. The top half of the cast-iron first shielding is hidden for clarity.}
    \label{fig:Upstream}

\end{figure}

The particle shower generated by the SPS beam impacting into the core could deposit up to $\sim$10~kW of power in the dump's vacuum chamber. To keep temperature gradients and the resulting thermal stresses below acceptable values to guarantee efficient and reliable operation, the chamber is actively cooled by forced air convection. The cooling process involves extracting air from the downstream side of the dump within the ECX5 cavern of the SPS tunnel, where TIDVG\#5 is installed. This air is then channeled into a 10-mm annular gap between the chamber and the external cast-iron shielding. Figure~\ref{fig:Upstream} illustrates this cooling scheme, highlighting that the airflow is primarily directed toward the two sides of the chamber, where the highest temperatures are experienced, and Table~\ref{tab:Table_6} summarizes the main characteristics of the air-cooling system.

\begin{table}[htb]
    \centering
    \caption{Main characteristics of air-cooling system.}
    \begin{tabular}{ C{5cm}  C{3cm} }
    \hline \hline
    Parameter & Value \\
    \hline
    Chamber outer diameter & 359~mm \\
    First shielding inner diameter & 379~mm \\
    Radial gap & 10~mm\\
    Average air velocity & $\sim$8~m/s\\
    Average volumetric flow & 500~m$^{3}$/h\\
    Inlet air temperature & 28$^{\circ}$C \\
    Outlet air temperature & $\sim$70$^{\circ}$C \\
    \hline \hline
    \end{tabular}
    \label{tab:Table_6}
\end{table}

\begin{figure}
    \centering\includegraphics[width=0.48\textwidth]{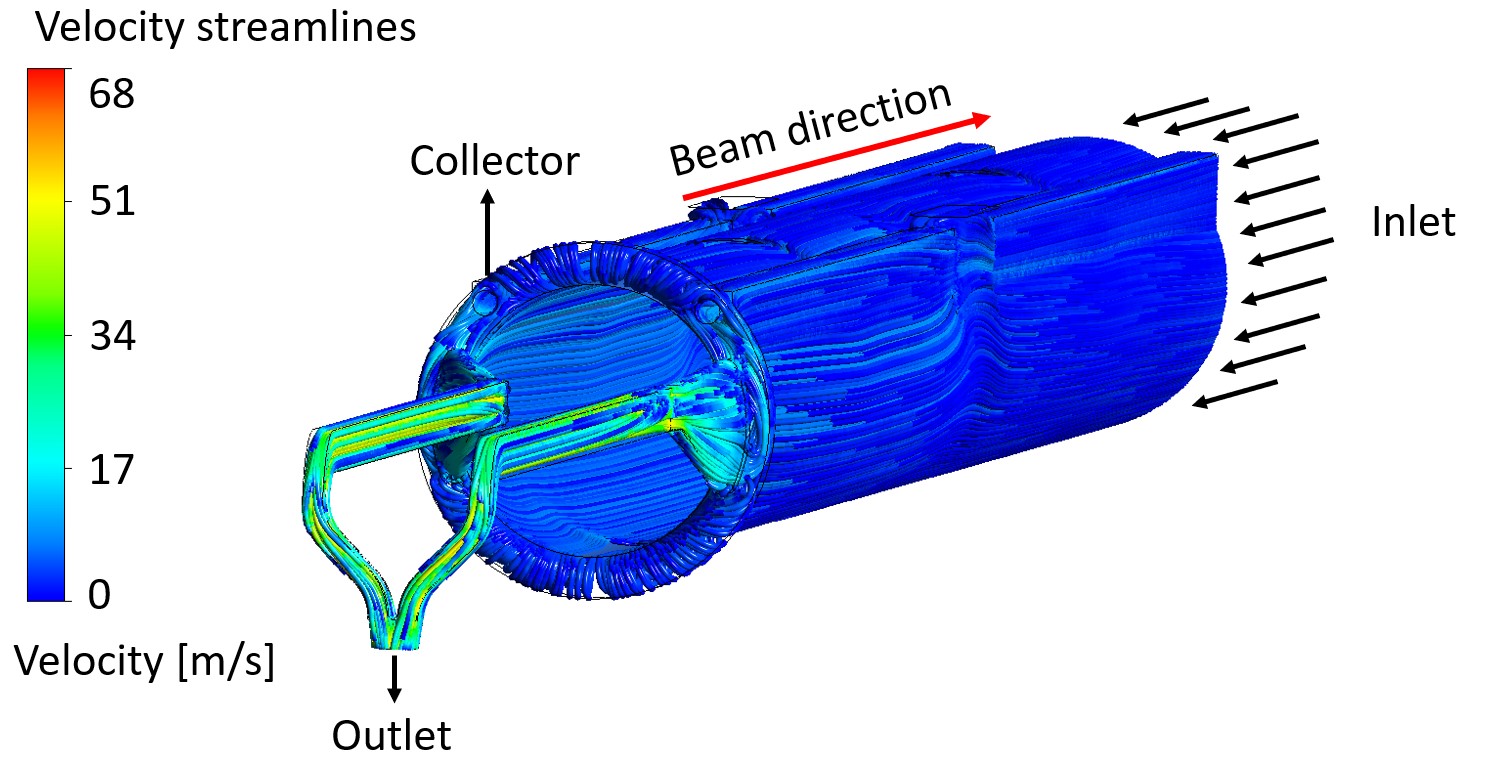}
    \caption{Velocity streamlines in the fluid domain of the air-cooling circuit around the vacuum chamber.}
    \label{fig:Streamlines}
\end{figure}

Based on the calculated velocity distribution (Fig.~\ref{fig:Streamlines}), the total pressure drop between the inlet and outlet of the dump's circuit is around $\sim$4250~Pa. Nevertheless, this value is significantly impacted by the design of the collector, where the most substantial pressure loss is observed, 1754~Pa. This is a consequence of the intricate design of this component, through which the airflow navigates complex geometries (Fig.~\ref{fig:Upstream}).

Given these flow characteristics, in the worst-case scenario, the average heat-transfer coefficient (HTC) at the chamber's walls is $\sim$50~W/(m$^{2}$\,K) and the maximum temperature is $\sim$168$^{\circ}$C. The required flow rate is maintained by two redundant 1000-m$^{3}$/h, 7000-Pa fans (only one operates at a time). A periodic switch between the two fans is implemented regularly during beam operation. It is intended that the air-cooling system will run continuously. However, the ventilation can work at a reduced flow during technical stops and/or maintenance operations, and it can be completely stopped if required. Furthermore, the air-cooling system contributes to creating sufficient under-pressure in the external multi-layered shielding, preventing the possible escape of activated dust during maintenance and operations involving opening the shielding.

\section{First and external multilayered shielding}

The TIDVG\#5 internal beam dump is one of the SPS's most radioactive pieces of equipment. In fact, the operation of the previous SPS dumps resulted in significant radiation-related issues. these can be categorized as follows:
\begin{enumerate}
\item Elevated residual dose rates and material activation: Activation of the beam dump caused high dose-rate levels in its immediate and surrounding areas, which are accessible during beam stop periods. After normal beam operation followed by a cool-down period of 30~h, the dose rate reached values in the range of 10~mSv/h at a lateral distance of approximately 70~cm from the dump. Even higher values were measured after dedicated heavy beam-dump operations.
\item Airborne radioactivity: Due to the beam dump being unshielded, high levels of airborne radioactivity were created.
\item Cable damage: The cascade of secondary particles emanating from the beam dump led to the deterioration of cable insulation located in close proximity to the dump.
\end{enumerate}
To mitigate these problems, the new TIDVG\#5 incorporates a robust shielding system comprising inner (first) and external shielding layers. The former encloses the core with two cylindrical blocks, while the latter is made from 40-cm-thick concrete followed by a 1-m layer of iron, capped with a 40-cm layer of concrete or marble. On the downstream side of the dump, two masks are positioned to capture high-energy particles that propagates along the beam line.

\subsection{First shielding}
\begin{figure}[ht]
    \centering
    \includegraphics[width=0.5\textwidth]{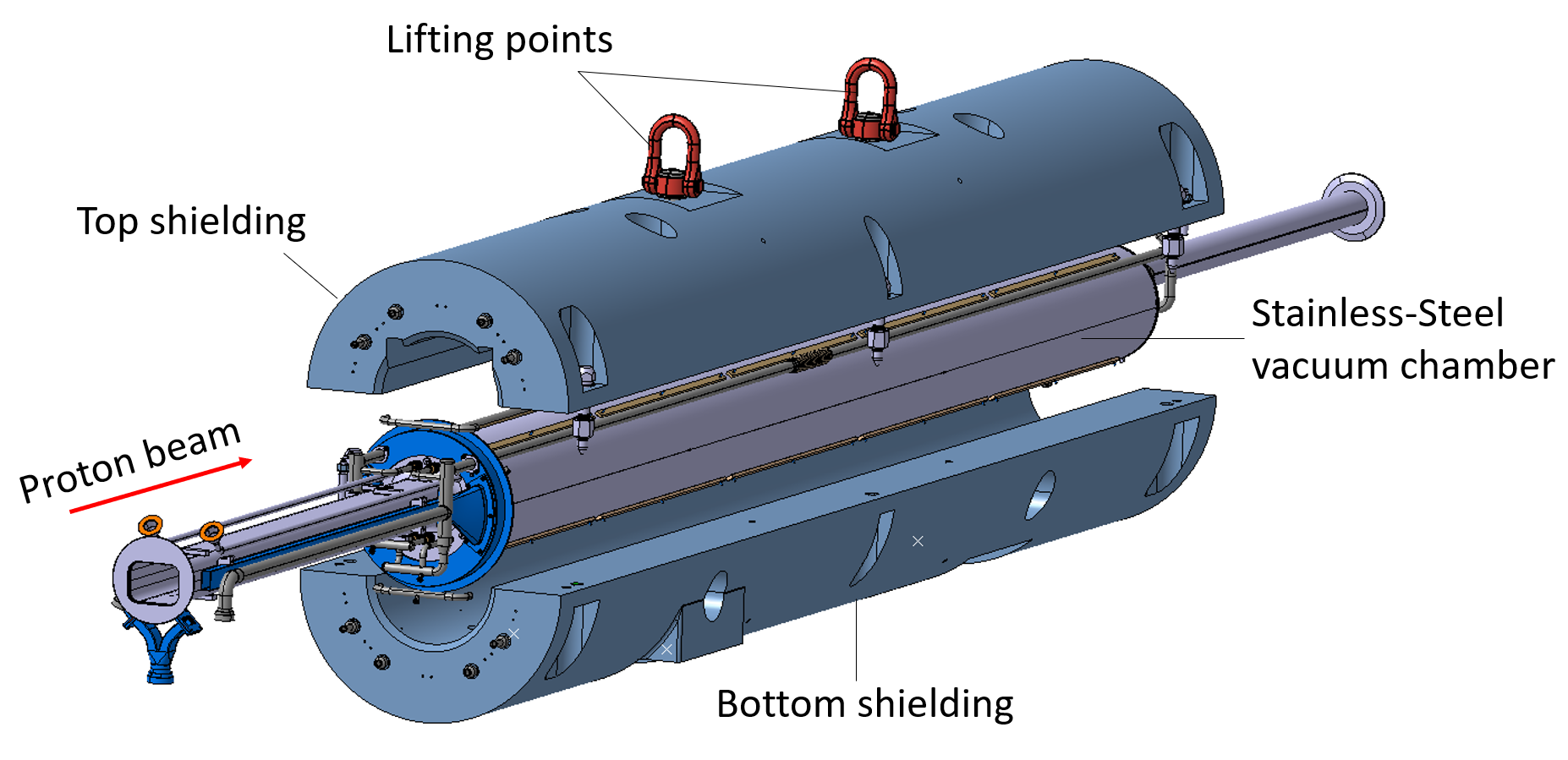}
    \caption{Exploded view of the top and bottom shielding with the inside core visible.}
    \label{fig:FS-Core}
\end{figure}

The first shielding encloses the core, as shown in Fig.~\ref{fig:FS-Core}, and it provides initial protection for nearby equipment from the radiation field generated while beam dumping. It is made from two cylindrical blocks of EN-GJS-400-18U-LT spheroidal-graphite cast iron according to EN~1563. This material was chosen for its particular ductility, which is the result of the spheroidal form of the type-V and -VI graphite, as described by the EN~1563 and EN~ISO~945 standards. The material of the first shielding being ductile is useful around the upper lifting points and the contact surface between the top and bottom shielding (Fig.~\ref{fig:FS-Core}), where large local stresses may be generated during handling. 

Table~\ref{tab:Tensile_Graphite} shows the measured tensile properties of this material at room temperature. Tight dimensional and shape tolerances---especially of the contact and inner surfaces---allow precise assembly of the blocks, minimizing the gaps between them and keeping the radiation leakage as low as possible. The blocks are actively cooled to dissipate the heat produced by the interaction of the bulk material with the beam shower. Four 316L SS cooling pipes per block, with inner and outer diameters of 16 and 36~mm, respectively, were embedded directly in the mold during the casting process to achieve good thermal-transfer efficiency from the blocks to the cooling water. To avoid thermal shocks and minimize deformations, the tubes were pre-heated to roughly $\sim$500$^{\circ}$C just before casting. Each block is equipped with remote-handling-friendly lifting points, allowing for remote installation and/or replacement operations.

\begin{table}[htb]
    \centering
    \caption{Measured tensile properties of the EN-GJS-400-18U-RT spheroidal-graphite cast iron at room temperature (20$^{\circ}$C).}
    \begin{tabular}{c c c}
    \hline \hline
     Yield stress (MPa) & Tensile strength (MPa) & Elongation (\%) \\
     \hline
         390 & 445 & 16 \\
    \hline \hline
    \end{tabular}
    \label{tab:Tensile_Graphite}
\end{table}

\subsection{External multilayered shielding}
The concept of the shielding design is to effectively minimize the propagation of the secondary particle cascade before it reaches areas that are frequently accessed during maintenance periods. This strategy reduces activation and therefore residual dose rates next to accessible parts of the beam dump. The rationale behind the incorporation of each individual layer is outlined as follows:
\begin{enumerate}
    \item Innermost concrete layer: when interventions are necessary at the dump core's location, especially during dump removal, employing an innermost layer composed of concrete results in significantly lower residual dose rates compared to an innermost iron layer. The latter would exhibit similar activation levels as the dump itself.
    \item Middle iron layer: the 1-m-thick iron layer, made of the same type of cast iron as that used for the first shielding (EN-GJS-400-18U-LT), reduces the secondary particle cascade by a factor of about 350. Due to activation processes, the iron shielding becomes radioactive, meaning it is a source of significant residual dose rate after beam operation. To counteract this side effect, a third shielding layer is positioned adjacent to the iron shielding.
    \item Outer concrete or marble layer: to effectively capture photons emanating from the activated iron layer, the shielding configuration ends with a 40-cm layer made of either concrete or marble, as shown in Figs.~\ref{fig:TIDVG_Assemb} and \ref{fig:TIDVG_Assemb1}. Compared to iron, this shielding layer is subject to lower activation production, and it also serves the purpose of capturing gamma radiation emitted by the highly activated iron layer. Furthermore, white marble plays a unique role among the various kinds of shielding materials due to its lower activation when compared to concrete thanks to its purity~\cite{radiation}. For this reason, to minimize the residual dose rate in the vicinity of the dump after short cool-down periods, it is used on the three sides accessible by personnel. To ensure its effectiveness, the chemical composition of the white marble employed was strictly controlled. Specifically, the minimum content of calcium carbonate (CaCO$_3$) had to be greater than 98\%, and all kinds of impurities were minimized (e.g. Na, Li, Co and Eu to less than 1.4\% and Sr less than 160 ppm).
\end{enumerate}

The attenuation of the high-energy radiation field provided by the shielding can be estimated as a factor of around 3000. This calculation, however, does not include the reduction in dose rates that is found next to accessible parts of the beam-dump system, which comes into play when comparing the previous dump concept with the newly implemented design.

\begin{figure}
    \centering
    \includegraphics[width=0.48\textwidth]{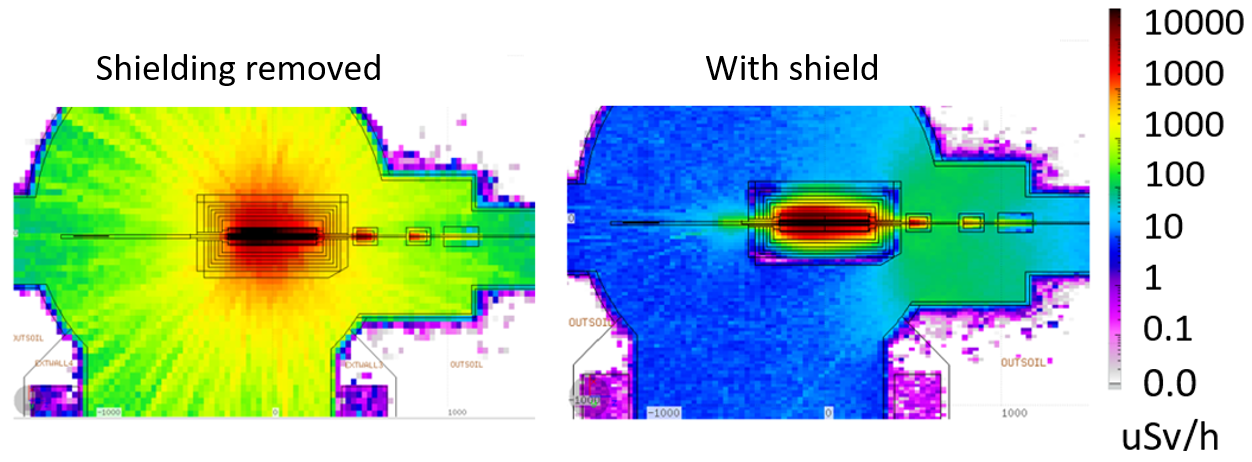}
    \caption{Comparison of residual dose rates, showing an unshielded dump core (on the left) and a shielded dump core (on the right). For both scenarios, the adopted beam-operation scheme, spanning 20~years of standard operation followed by a single day of intense dump usage, is followed by a 1-week cool-down period. }
    \label{fig:shield}
    
\end{figure}

In addition to the shielding structure, two masks are positioned downstream of the dump configuration to intercept the forward-directed high-energy particle cascade that arises from the impact of the primary beam on the dump core. These masks, which are designed to capture high-energy particles, consist of a structural steel (S355) core enveloped by a layer of marble. Figure~\ref{fig:shield} shows a comparison of residual-dose-rate scenarios, contrasting an unshielded condition (in which shielding components are filled with air) with the shielded dump-core configuration. This simulation considered a beam-operation schedule involving 20~years of standard operation followed by a single day of intense dump usage. Subsequent to beam operation, there is a cool-down period of 1~week. When analyzing the lateral accessible regions and comparing the shielded and unshielded dump conditions, a significant dose-rate difference of a maximum factor of 10\,000 emerges. This dose-rate disparity between the two scenarios is consistent with the previously mentioned attenuation effect triggered by the shielding (a factor of approximately 3000), the increased distance from the beam line in the shielded configuration, and the different material types adjacent to the accessible areas.

The implemented shielding setup yields additional beneficial outcomes compared to previous versions; notably, these include a reduction in cable-insulation doses and diminished air activation external to the shielding. Both these advantages can be traced back to the radiation-attenuation power of the shielding enveloping the dump core. Most of the air activation occurs within the beam-dump shielding and along the beam axis in the forward direction from the dump. To minimize air activation within the shielding adjacent to the dump core, efforts were made to minimize the air volume in this specific region.

\begin{figure}
    \centering\includegraphics[width=0.48\textwidth]{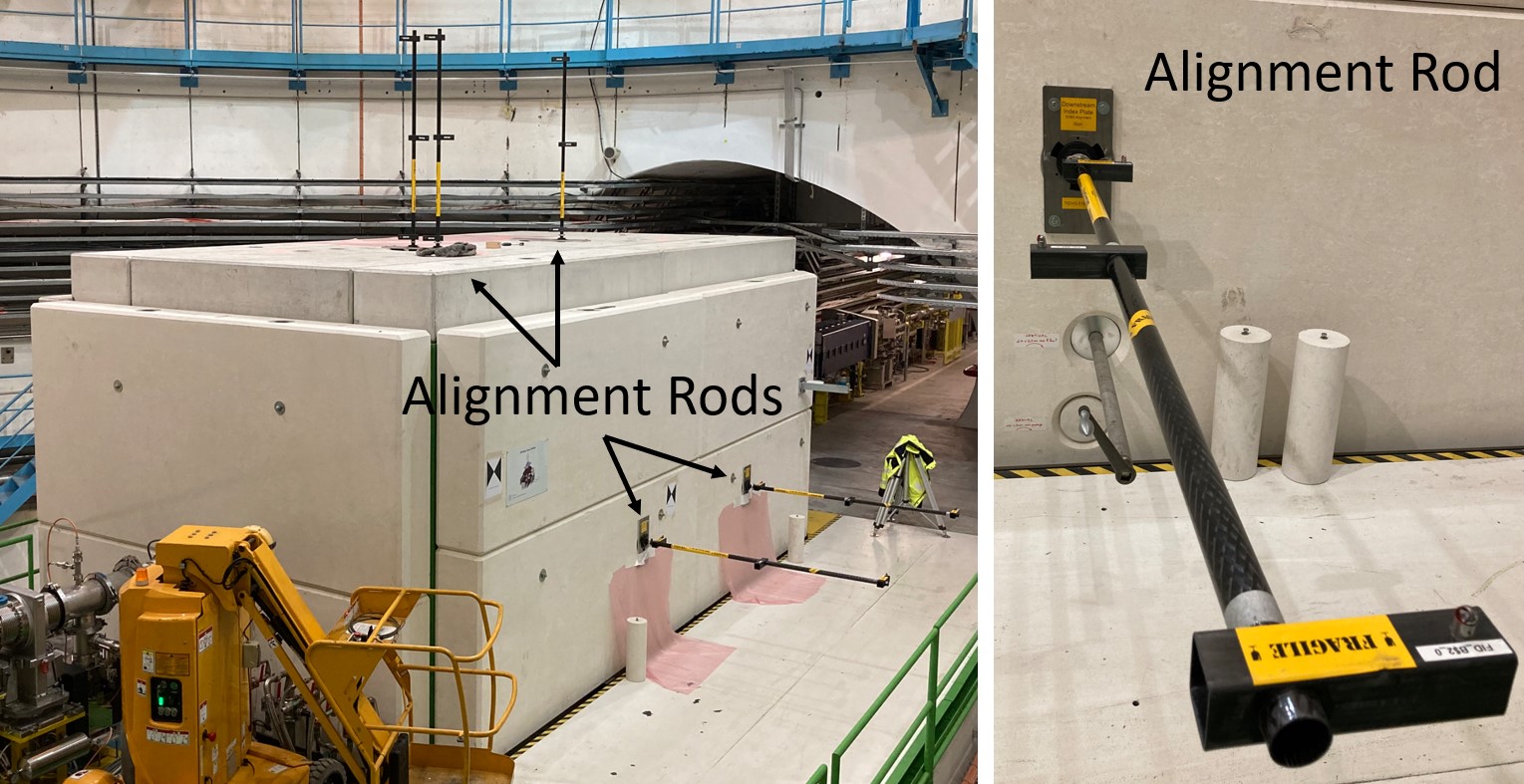}
    \caption{Alignment rods installed during inspection.}
    \label{align}
\end{figure}

The design objective for environmental air release applied to the new dump aligns well with CERN's optimization criterion, which stipulates that the annual dose received by any member of the public due to CERN's operations should not exceed 10~\textmu Sv. To validate the compliance with this criterion, two independent measurement campaigns were executed to assess the airborne radioactivity released. Both campaigns projected a release of $9.1 \times 10^{12}$~Bq per year, accounting for $2 \times 10^{18}$ dumped protons annually. The difference between the two measurements fell within a 2\% range. The resulting airborne radioactivity release corresponds to an annual committed dose received by the reference group of the public of 1.6~\textmu Sv. This value is less than 1\% of the annual dose received in the Geneva area due to natural causes. Further details of the FLUKA simulation studies concerning the design of the beam-dump shielding and the radiological impact of the dump's operation can be found in Refs.~\cite{RP} and~\cite{residual}. In case replacement of the dump is required, the different blocks are arranged so as to minimize the handling operations required. Moreover, similarly to the cast-iron first shielding, every block has been designed to be handled remotely. Both the first and external shielding present service through-holes to allow for survey and alignment operations of the dump whilst inside the shielding (Fig.~\ref{align}).

The dump and its first shielding are resting on three jack assemblies that allow to adjust two linear degrees of freedom each: one vertical and one horizontal. These components have been designed specifically for this application and have been tested to comply with requirements.
Through the array of these three supports, the dump can be precisely aligned in all six degrees of freedom. However, due to the proximity to the beam, the jacks are expected to be exposed to a significant dose over their lifetime (20 years), which can be estimated to be as high as 10 MGy.
Based on this, the materials and lubricants used for these assemblies are the result of a careful selection process, taking into account their behaviour under this type of irradiation~\cite{FERRARI2021101088}.

\section{Numerical studies}
\subsection{Beam parameters and most demanding super-cycle}
During its lifetime, the TIDVG\#5 will operate with various beam types distributed in complex predefined sequences known as super-cycles. Table~\ref{tab2} shows two beam types (cycle and energy per pulse), which serve as the basis for the numerical analysis presented in this section. The combination of these two super-cycles yields two critical scenarios: the nominal case ($\sim$164~kW) and the worst-case operational situation ($\sim$270~kW), which the TIDVG\#5 is specifically designed to withstand.

\begin{table}[ht]
    \centering
    \caption{Beam types analyzed for steady-state cases.}
    \begin{tabular}{C{4cm} C{4cm} } 
        \toprule \toprule
        Beam cycle & Energy per pulse (MJ) \\
        \midrule
        LIU-SPS~80b& 5.60  \\
        SPS-FT SHiP &  2.88  \\
        \bottomrule \bottomrule
    \end{tabular}
    \label{tab2}
\end{table}

The first super-cycle, referred to as LHC Filling, corresponds to nominal operation and consists in a combination of one LIU-SPS~80b pulse followed by an SPS-FT SHiP pulse, in a super-cycle period of 36 seconds. This results in an average beam power of $\sim$164~kW. The worst-case scenario, referred to as FT Production, involves the sequence of two consecutive SPS-FT SHiP pulses in a super-cycle period of 10.8 seconds, for a total average beam power of $\sim$270~kW.

\subsection{Beam-energy deposition}
\begin{figure}[htb]
    \centering
    \includegraphics[width=0.48\textwidth]{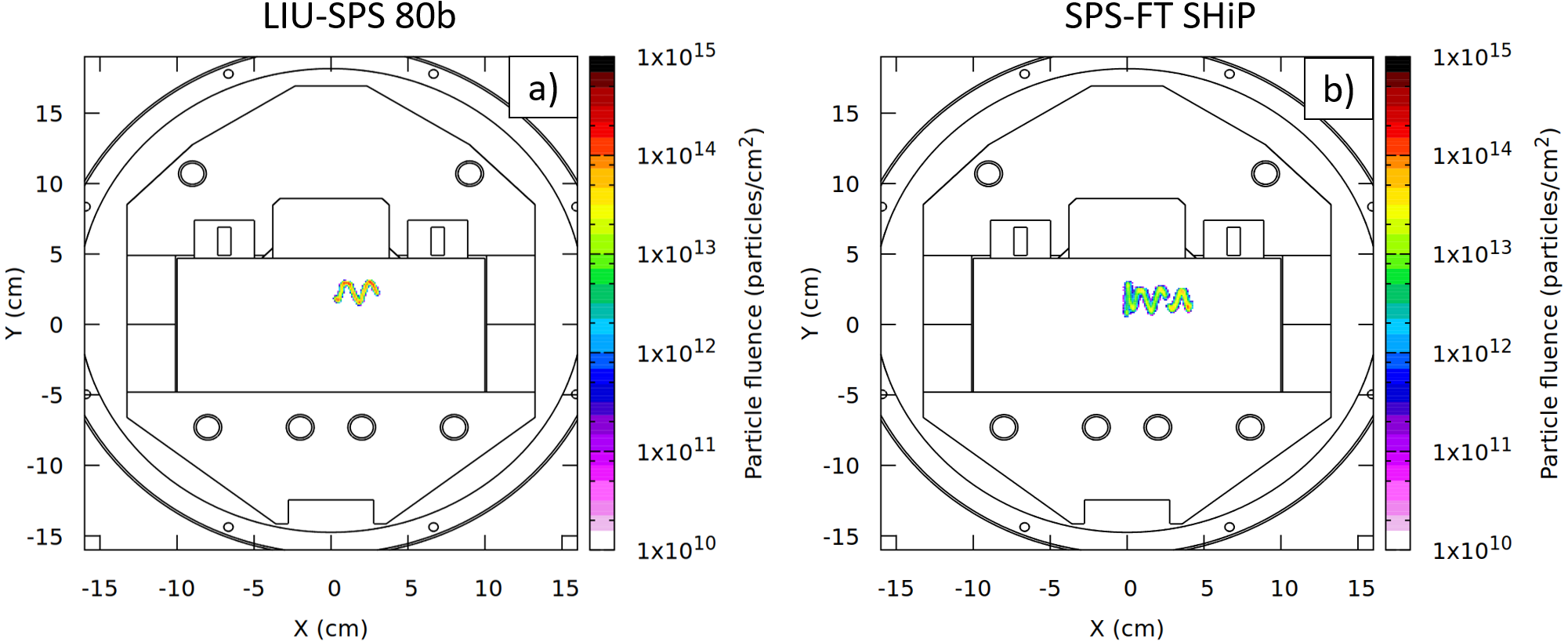}
    \caption{Beam profiles for the LIU-SPS~80b (left) and SPS-FT SHiP (right) beams, viewed from downstream.}
    \label{fig:deposition}
\end{figure}

The vertical and horizontal kicker magnets deflect the beam toward the absorbing blocks, leaving a sinusoidal pattern, as shown in Figs.~\ref{fig:deposition}(a) and \ref{fig:deposition}(b). In both scenarios, the pattern is asymmetrical with respect to the center of the dump.

\begin{figure}[htb]
    \centering
    \includegraphics[width=0.48\textwidth]{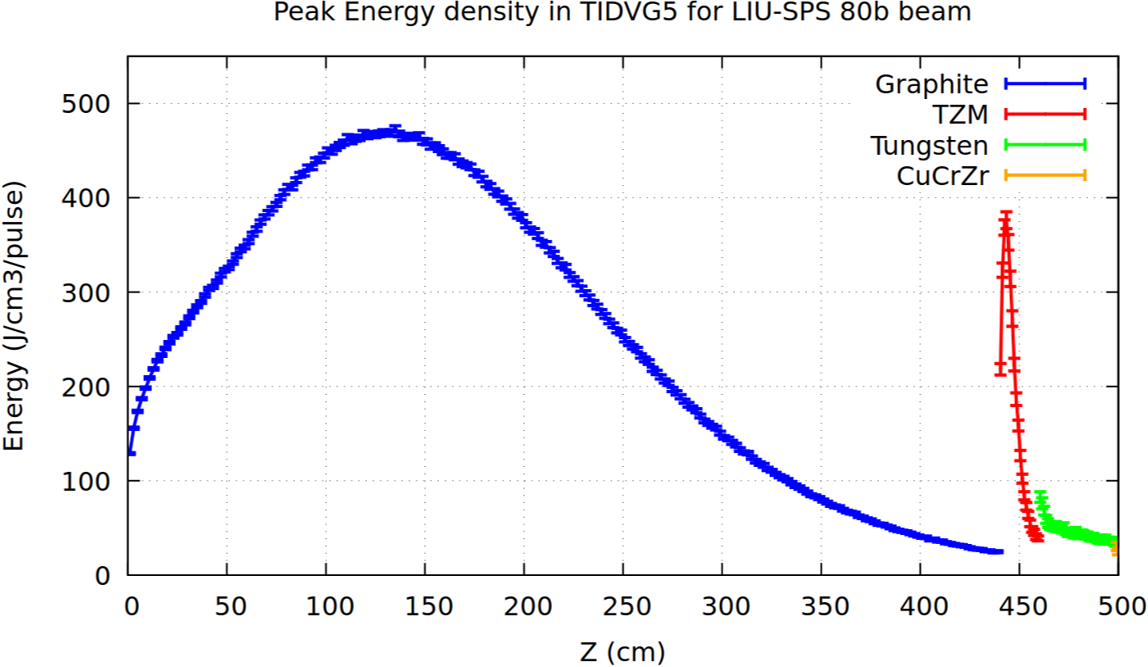}
    \caption{Peak energy density deposited along the length of the dump for a pulse of LIU-SPS 80b, as calculated with the FLUKA Monte Carlo code~\cite{10.3389/fphy.2021.788253}.}
    \label{fig:deposit}
\end{figure}

The FLUKA Monte Carlo code is employed for all beam-matter and radiation protection calculations ~\cite{10.3389/fphy.2021.788253}. As is evident from the plot presented in Fig.~\ref{fig:deposit}, the maximum energy deposition occurs within the fourth graphite block. Additionally, the TZM and tungsten blocks are strategically positioned to attenuate the residual energy of the beam, serving as a protective barrier for downstream devices.

\subsection{Thermomechanical studies of the absorbing blocks}
The absorbing blocks are positioned atop the CuCr1Zr cooling plates; to facilitate unimpeded expansion, these are pressed together without applying excessive force. Consequently, heat dissipation occurs primarily through conduction within the blocks; this is followed by transfer to the cooling plates, during which the TCC value assumes a pivotal role in enhancing the efficiency of heat dissipation.

Drawing from previous experience with the TIDVG\#4, identical values for the TCC coefficients were initially assumed for the contact between graphite and the CuCr1Zr cooling plates. For the contact between tungsten and CuCr1Zr, the same TCC value as for TIDVG\#4 was assumend, even if the TIDVG\#4 used Inermet180, a tungsten-based alloy. Finally, the contact between TZM and CuCr1Zr was chosen to be equal to that of tungsten; which is a conservative approach because the latter has a higher Young's modulus and the contact conductance between TZM and CuCr1Zr should thus be better. All of these coefficients underwent adjustment to align with operational feedback for TIDVG\#5, refining their accuracy, and they are listed in Table~\ref{TCCTable}. It is worth noting that all the results presented in this section are based on the use of the TIDVG\#4-tuned values for the new version.

\begin{table}[h!t]
    \centering
    \caption{TCC values between system components used for thermomechanical studies of the absorbing blocks.}
    \begin{tabular}{C{3cm} C{2.5cm} C{2.5cm}}
    \hline \hline
           & TIDVG\#4-tuned TCC [W/(m$^{2}$\,K)] & TIDVG\#5-tuned TCC [W/(m$^{2}$\,K)]\\
         \hline
          Graphite--CuCr1Zr cooling plates & 2000 & 900 \\
          TZM--CuCr1Zr cooling plates & 800 & 900 \\
          Tungsten--CuCr1Zr cooling plates & 800 & 800 \\
         \hline \hline
    \end{tabular}
    \label{TCCTable}
\end{table}

As revealed by the operational feedback from TIDVG\#5, there is a notable decrease in the TCC between the graphite blocks and CuCr1Zr cooling plates compared to the previous version. Multiple uncontrolled variables could potentially influence this value, including irregular surface roughness, disparities in flatness, interactions between adjacent blocks stemming from thermal expansion, and other related factors.

As previously noted, the TIDVG\#5 absorbing blocks are designed to withstand an average beam power of $\sim$270~kW. Therefore, to assess the thermomechanical response of the blocks, FEM simulations were carried out considering the loads they experience under a thermal steady state at 270~kW. To achieve this scenario, the absorbing blocks must be intercepting SPS-FT SHiP beams for a duration of 50~min. Table~\ref{tab3} lists the peak temperatures achieved in the absorbing blocks under these conditions.

\begin{table}[h!t]
    \centering
    \caption{Peak temperatures in the absorbing blocks under steady-state conditions at 270~kW. }
    \begin{tabular}{C{4cm} C{4cm}} 
        \toprule \toprule
        Absorbing block & Peak temperature ($^{\circ}$C) \\
        \midrule
        Graphite & 532 \\
        TZM & 505 \\
        Tungsten & 98 \\
        \bottomrule \bottomrule
    \end{tabular}
    \label{tab3}
\end{table}

Temperature gradients can lead to large thermal stresses, and these are determined by uniform or nonuniform temperature changes in a body that is somehow constrained against expansion or contraction~\cite{thermalloads}. In the presence of temperature gradients, the existence of such constraints will lead to the exertion of either tensile or compressive forces in the absorbing blocks. The structural integrity of a material can be assessed by means of either the von~Mises or Christensen failure criteria~\cite{mises,Christensen}, depending on whether the material is ductile or brittle. When the criterion parameter reaches a value of ``1,'' a brittle specimen is likely to break, while a ductile specimen is expected to enter the plastic domain. Table~\ref{tab4} shows the results for the most critical absorbing blocks per material.

\begin{table}[h!t]
    \centering
    \caption{Christensen and von~Mises failure criterion values for the absorbing blocks. }
    \begin{tabular}{C{4cm} C{4cm}} 
        \toprule \toprule
        Absorbing block & Criterion $<$ 1 \\
        \midrule
        Graphite (Christensen) & 0.35    \\
        TZM (von~Mises) & 0.98 \\
        Tungsten (Christensen) & 0.37 \\
        \bottomrule \bottomrule
    \end{tabular}
    \label{tab4}
\end{table}

Based on these failure criteria, it can be seen that none of the absorbing blocks exceeds the allowable limit. Both the graphite and tungsten blocks (for which the Christensen criterion was used due to their brittle nature) are expected to be able to withstand the worst-case scenario. Nonetheless, in the case of the first TZM block (for which the von~Mises criterion is applied due to its ductile behavior at high temperatures), the failure formula reveals a borderline value.

While the von~Mises stress in the TZM block approaches the critical value at its most stress-prone location, it is crucial to understand that this does not automatically imply failure, especially for ductile materials. Should the limiting threshold be surpassed, TZM, which is known for its ductile behavior at elevated temperatures, would experience strain hardening at the point of beam impact and continue to effectively absorb and dissipate proton beams.

\subsection{Thermomechanical FEM studies of the CuCr1Zr core and vacuum chamber}
FEM models of the CuCr1Zr cooling plates and the vacuum chamber were studied to predict the strain levels induced within these components by the impact of the particle shower. In contrast to the absorbing blocks, the CuCr1Zr core and the vacuum chamber may potentially experience some plastic deformation in a worst-case scenario. To provide a comprehensive assessment, both components were also analyzed under more typical operating conditions, specifically, during steady-state operation with LHC Filling super-cycles. This approach allows to gain a comprehensive understanding of the thermomechanical behavior within the dump system.

\begin{figure}[htb]
    \centering
    \includegraphics[width=0.48\textwidth]{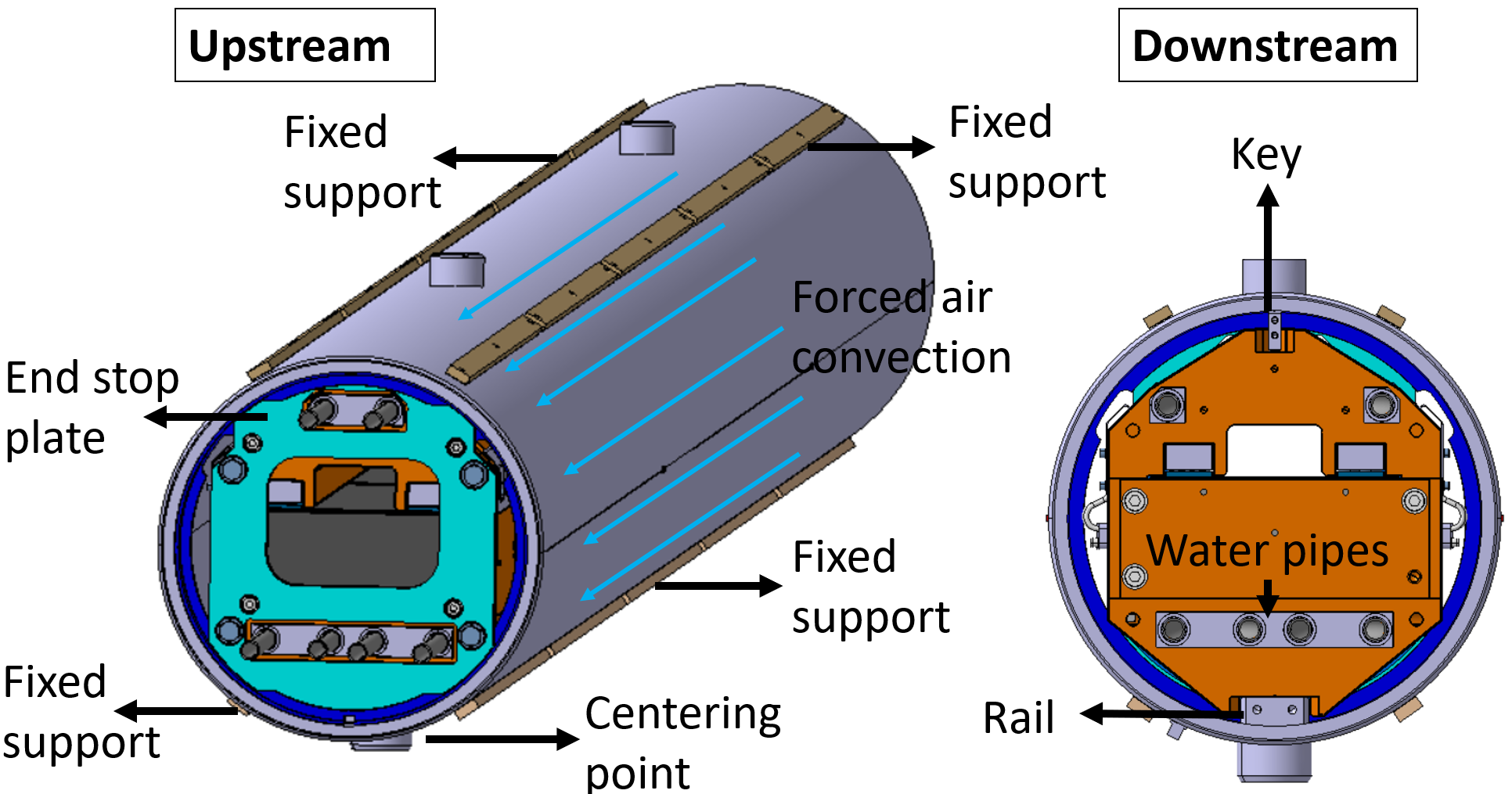}
    \caption{Schematic representation of contacts and constrains in (a)~downstream and (b)~upstream views.}
    \label{Contacts}
\end{figure}

Figure~\ref{Contacts} illustrates the boundary conditions applied to the models of the vacuum chamber and the inner CuCr1Zr plates. The TIDVG\#5 core is supported within the cast-iron first shielding by the bottom and top supports. The beam aperture of the dump is the smallest of the SPS, and it has to stay within precise tolerances in accordance with the theoretical beam axis. The top/bottom supports help to limit the dump's vertical displacement during operation and maintain these tolerances. The centering point locates the whole assembly within the first shielding. The inner CuCr1Zr cooling plates and absorbing blocks are resting on the bottom rail, and they are guided horizontally by the downstream top key and screwed onto the end-stop plate. The latter is mounted within a guide that is welded to the chamber.

\subsubsection{Static simulations}
\begin{figure}[htb]
    \centering
    \includegraphics[width=0.48\textwidth]{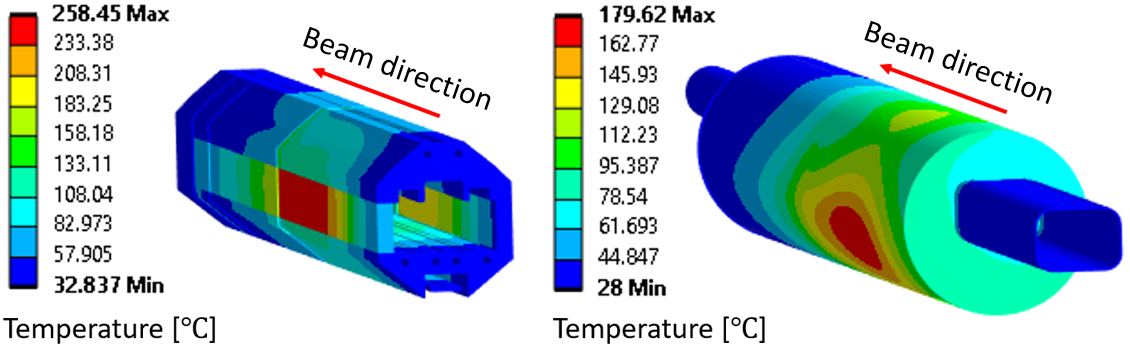}
    \caption{FT Production ($\sim$270~kW) FEM thermal simulations: (a)~CuCr1Zr core; (b)~vacuum chamber.}
    \label{therm270}
\end{figure}

\begin{figure}[htb]
    \centering
    \includegraphics[width=0.48\textwidth]{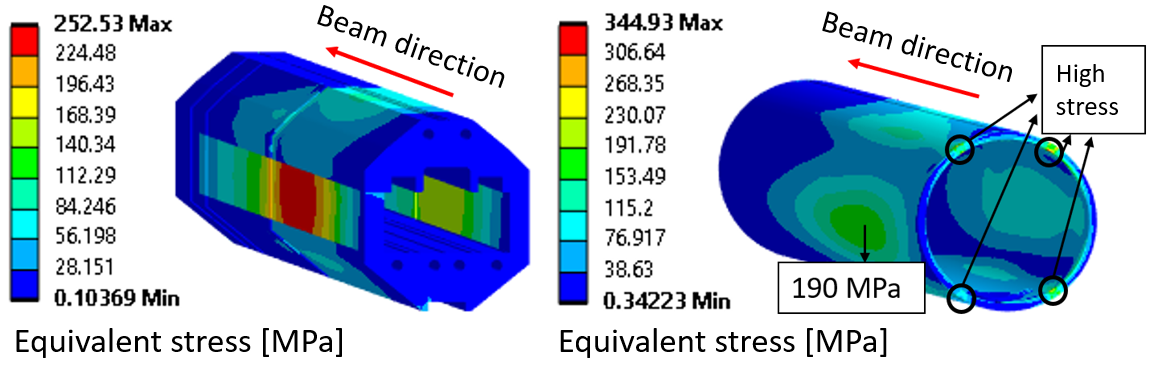}
    \caption{FT Production ($\sim$270~kW) FEM structural simulations: (a)~CuCr1Zr core; (b)~vacuum chamber.}
    \label{struct270}
\end{figure}

\textit{(a)~Worst-case scenario}. Given the thermal steady-state conditions at $\sim$270-kW average beam power, the temperature distribution in the CuCr1Zr core and the vacuum chamber are as shown in Figs.~\ref{therm270} and \ref{struct270}.

As can be observed in Fig.~\ref{therm270}(a), the temperatures within the CuCr1Zr core remain within reasonable limits for such a material. It is important to note that a significant reduction in thermophysical and mechanical properties typically occurs at temperatures exceeding 400$^{\circ}$C. With a peak temperature of 252$^{\circ}$C, there is no risk of annealing the material. The SS chamber depicted in Fig.~\ref{therm270}(b) also maintains temperatures within acceptable ranges for the material, under the assumption that this represents a worst-case scenario.

Nonetheless, it is crucial to consider the thermal gradients induced by these conditions; as noted, thermal gradients can result in high mechanical stresses that may lead to strain hardening of the material. In the case of the CuCr1Zr core (Fig.~\ref{struct270}), the occurrence of plastic deformation is highly unlikely. If it were to occur, its impact would likely be negligible. However, the vacuum chamber shown in Fig.~\ref{struct270}(b) is subjected to high stress concentrations at its extremities. This is primarily due to the assumption that the chamber walls are bonded to the rest of the structure, representing a conservative contact condition. Furthermore, the equivalent stresses detected in the chamber walls suggest the possibility of plastic deformation.

\begin{figure}[htb]
    \centering
    \includegraphics[width=0.48\textwidth]{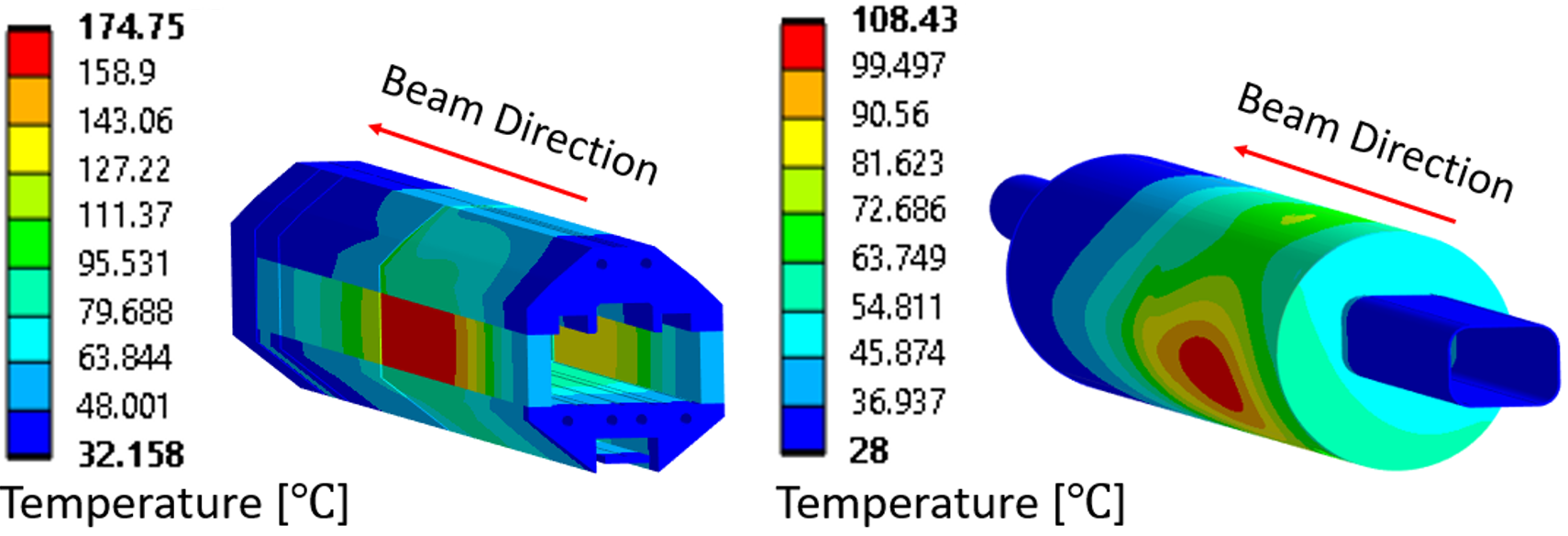}
    \caption{Results of LHC Filling steady-state FEM thermal simulations: (a)~CuCr1Zr cooling plates; (b)~vacuum chamber.}
    \label{therm}
\end{figure}

\textit{(b)~Nominal operational scenario}. The following calculations consider the steady-state conditions created by LHC Filling super-cycles and an average beam thermal power of 164~kW. Figure~\ref{therm} depicts the corresponding temperature distributions on the CuCr1Zr core and vacuum chamber.

\begin{figure}[htb]
    \centering
    \includegraphics[width=0.48\textwidth]{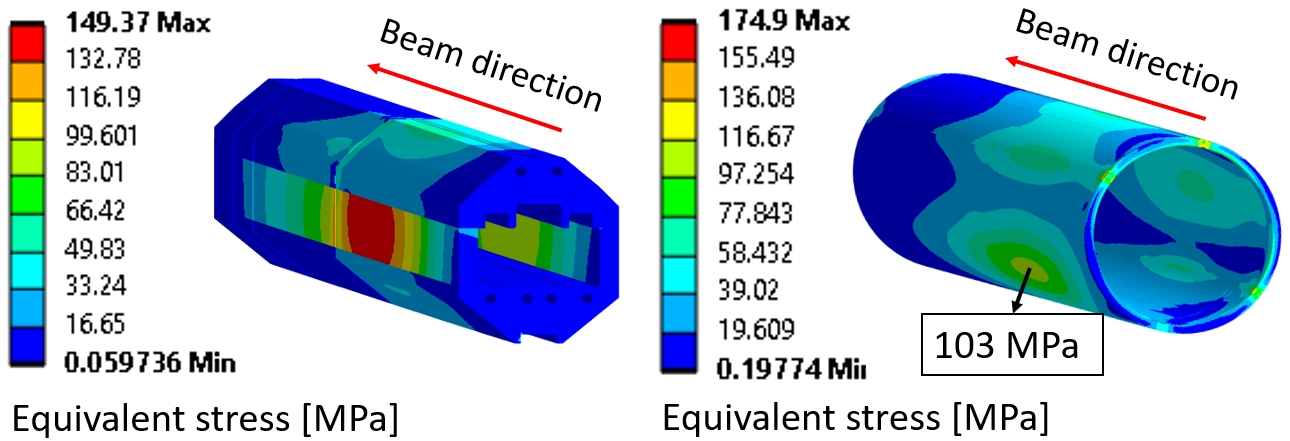}
    \caption{Results of LHC Filling steady-state FEM structural simulations: (a)~CuCr1Zr cooling plates; (b)~vacuum chamber.}
    \label{struct}
\end{figure}

The thermal distribution observed in Fig.~\ref{therm} highlights the regions of the cooling plates and vacuum chamber where the maximum temperatures are reached, i.e. on the left-hand side wall with respect to the particle-beam direction; the peak values are 175$^{\circ}$C and 108$^{\circ}$C, respectively.

Figures~\ref{struct}(a) and \ref{struct}(b) show large von~Mises stress concentrations in the CuCr1Zr cooling plates ($\sim$150~MPa) and vacuum chamber (103~MPa) located in the same areas in which the peak temperatures are achieved in Fig.~\ref{therm}. However, the maximum von~Mises stresses in the vacuum chamber, as can be observed in Fig.~\ref{struct}(b), are located at the interface with its top supports. When these stresses are compared to the yield-strength values listed in Table~\ref{yield2}, it can be seen that there are large safety margins, meaning that the materials are not expected to enter the plastic domain during this critical operational loading case.

\begin{table}[h!t]
    \centering
    \caption{Yield strengths of CuCr1Zr and 304L SS~\cite{Copper,304L} at different temperatures. The 304L SS yield-strength values at high temperatures are approximated. }
    \begin{tabular}{C{3cm} C{2cm} C{2cm}}
    \hline \hline
         Temperature ($^{\circ}$C) & Yield strength of CuCr1Zr (MPa) & Yield strength of 304L SS (MPa)\\
         \hline
          20 & 280 & 200 \\
          100 & 275 & 150 \\
         200 & 260 & 118 \\
         250 & 250 &  110 \\
         300 & 230 & 100 \\
         \hline \hline
    \end{tabular}
    \label{yield2}
\end{table}

\subsubsection{Transient simulations}
According to its functional specification~\cite{beam}, the TIDVG\#5 SPS internal beam dump should be able to accept five consecutive high-energy, full intensity, FT SHiP beam dumps starting from steady-state conditions of LHC Filling super-cycles. To assess the thermal and structural integrity of the dump, this scenario was simulated.

\begin{figure}[htb]
    \centering
    \includegraphics[width=0.48\textwidth]{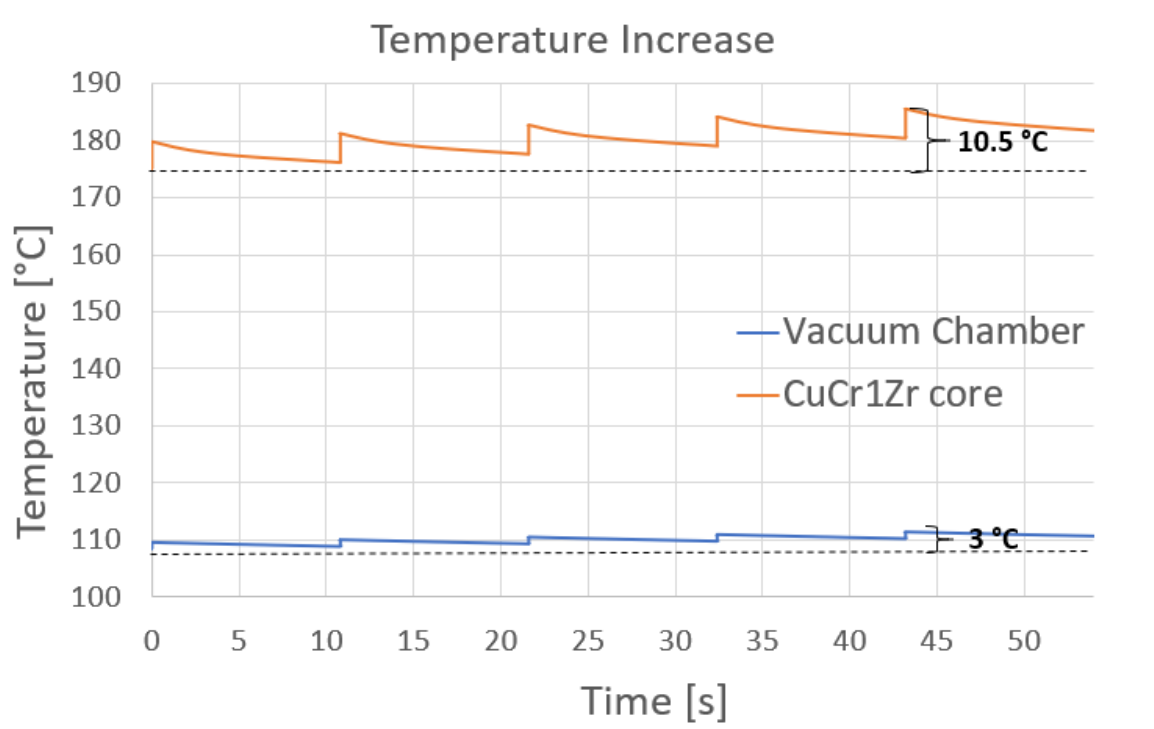}
    \caption{Results of transient FEM thermal simulations, showing the maximum temperature increases after five FT-SHiP pulses following steady-state LHC Filling.}
    \label{temperature}
\end{figure}

\begin{figure}[htb]
    \centering
    \includegraphics[width=0.48\textwidth]{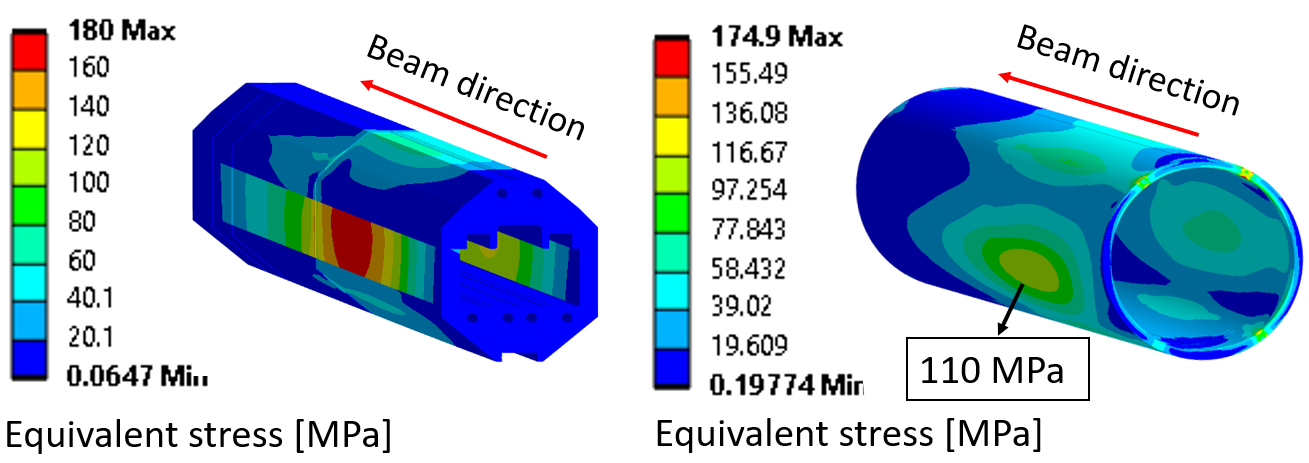}
    \caption{Results of transient FEM structural simulations after five FT-SHiP pulses following steady-state LHC Filling: (a)~CuCr1Zr core; (b)~vacuum chamber.}
    \label{transientstruct}
\end{figure}

The maximum temperature increases for the CuCr1Zr cooling plates and the vacuum chamber were found to be 10.5$^{\circ}$C and 3.0$^{\circ}$C, respectively, at the end of the fifth pulse (see Fig.~\ref{temperature}). The temperature change in the CuCr1Zr side plate has a non-negligible effect on its stresses; in contrast, stress relaxation predominates in the vacuum chamber, and negligible stress differences were found with respect to the initial LHC Filling steady-state conditions, as shown in Figs.~\ref{struct}(b) and \ref{transientstruct}(b).

Figure~\ref{transientstruct}(a) illustrates the increase in the peak von~Mises stress for the CuCr1Zr side plate with respect to the LHC Filling steady-state condition shown in Fig.~\ref{struct}(a). Despite this increase, all CuCr1Zr cooling plates are still expected to remain in the elastic domain.

\section{Instrumentation}
To monitor the actual thermomechanical behavior during operation, several sensors were installed in the TIDVG\#5 SPS internal beam dump, and these are detailed in this section.

\subsection{Temperature sensors, flowmeters, and LVTDs}

TIDVG\#5 is equipped with 35 Pt100-type temperature probes. These are distributed as follows. a) 24 on the absorbing blocks, b) two on the hottest points of the CuCr1Zr side plates, c) four on the vacuum chamber, d) two on the cast iron first shielding and e) three at the outlet of the first shielding and CuCr1Zr water-cooling circuits.

\begin{figure}[htb]
    \centering
    \includegraphics[width=0.48\textwidth]{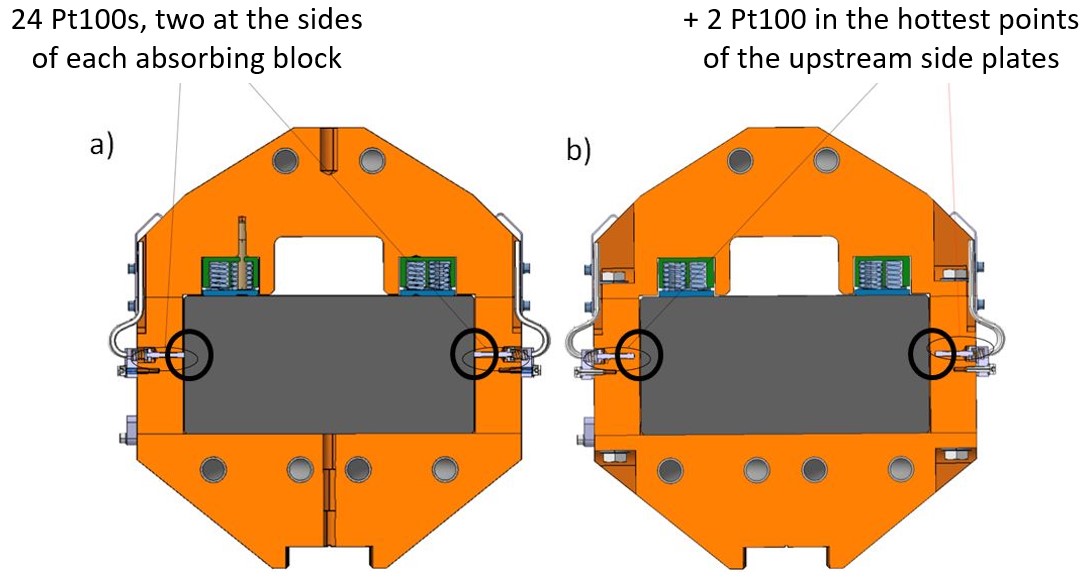}
    \caption{Position of the temperature sensors on: (a)~absorbing blocks; (b)~CuCr1Zr side plates.}
    \label{tempsensor}
\end{figure}

Each absorbing block is monitored via two Pt100 sensors on each side, left and right with respect to the particle-beam direction (24 sensors in total). The CuCr1Zr is monitored by only two Pt100s, which are located at the theoretical hottest points of the two upstream side plates. Both configurations are illustrated in Fig.~\ref{tempsensor}.

In the vacuum chamber, the temperature distribution at the time of operation has to be controlled to avoid excessive thermal stresses and displacements. Four temperature sensors are installed on the vacuum chamber: on the left and right of its upstream and downstream sides with respect to the beam direction. The upstream sensors will record the maximum temperatures, and the downstream sensors will provide information about the temperature gradients.

Finally, two temperature sensors are mounted at the inner diameter of the cast-iron first shielding, at approximately the same longitudinal positions as those installed on the vacuum chamber. For the cooling circuits, the three Pt100 sensors installed at the outlets of the CuCr1Zr and the top and bottom first-shielding circuits will be used to compute the thermal power extracted, along with the inlet temperatures and the flow rates.

The flow rates of the top/bottom first shielding and the CuCr1Zr cooling circuits are measured by two flow meters installed at the corresponding outlets. The air-flow rate and the temperature are also measured by a flow meter installed immediately at the outlet of the air-cooling system.

\begin{figure}[htb]
    \centering
    \includegraphics[width=0.45\textwidth]{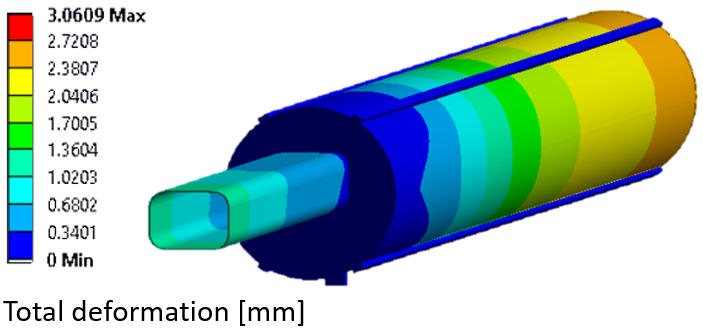}
    \caption{Total deformation of vacuum chamber in the operational scenario at 166~kW.}
    \label{deformation}
\end{figure}

The particle beam deposits energy on the dump's components asymmetrically. With respect to the beam direction, the left side is more heavily loaded then the right side. This results in non-negligible bending and hence displacements of the whole system. Figure~\ref{deformation} shows the expected total displacements of the dump's upstream and downstream ends when thermal energy is absorbed at a rate of 166~kW, accounting to up to 3~mm.

\begin{figure}[htb]
    \centering
    \includegraphics[width=0.48\textwidth]{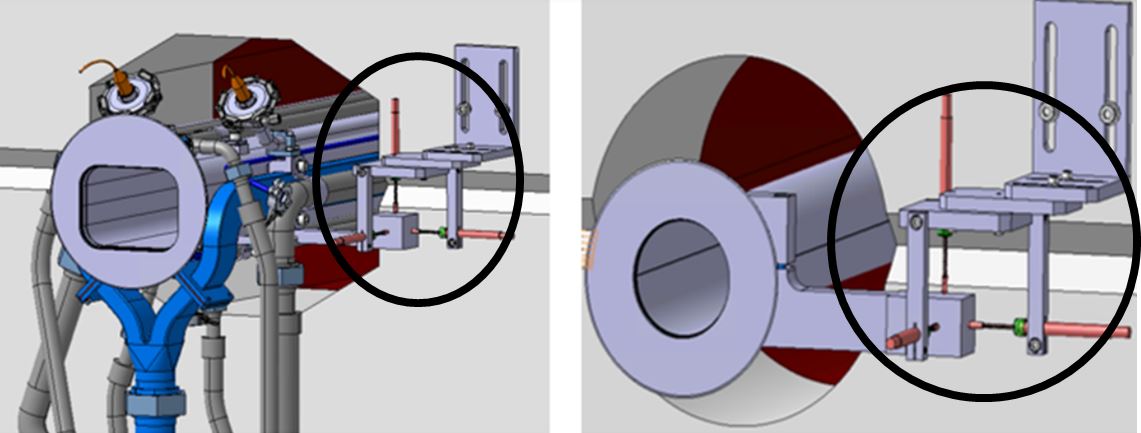}
    \caption{LVDT sensors: (a)~upstream; (b)~downstream.}
    \label{LVDTS}
\end{figure}

To compensate for these displacements, bellows are installed on either side of TIDVG\#5, and these will be monitored by six LVDTs in total, three upstream and three downstream, one for each direction, as shown in Fig.~\ref{LVDTS}. LVDTs are used to convert mechanical motion into a variable electrical current, voltage, or electrical signal.

\section{Operational feedback}
As noted earlier, TIDVG\#5 is designed to cope with an average beam thermal power of 166~kW in nominal operation and 270~kW in a worst-case scenario, both in steady-state. Nevertheless, at the time of writing, these conditions have not yet been achieved. 

\subsection{Commissioning beam parameters}
Since the beginning of its commissioning phase, TIDVG\#5~has reached thermal steady state on several occasions. Three scenarios with different average beam thermal powers were selected to be compared with the results of FEM simulations. The criteria to select a period of time to be compared were a) uniform initial temperature of all components, b) the average beam thermal power deposited in the blocks is constant in time and c) the selected period of time has to be sufficiently long for the all the components (absorbing blocks, CuCr1Zr core, and vacuum chamber) to reach thermal steady state.

Table~\ref{HRMT} lists three occasions on which such scenarios occurred, along with information regarding the average beam thermal power. Although beams of higher average thermal power have been intercepted by TIDVG\#5, these situations did not fulfill the requirements to reach steady state.

\begin{table}[ht]
    \centering
    \caption{Steady-state cases analyzed.}
    \begin{tabular}{c c} 
        \toprule \toprule
        Date & Average power (kW) \\
        \midrule
        26-06-2021 & 25 \\
        04-07-2021 & 41 \\
        29-03-2023 & 100 \\
        \bottomrule \bottomrule
    \end{tabular}
    \label{HRMT}
\end{table}

Although beams with higher average thermal power were intercepted near operational and worst-case scenario regimes, they failed to reach the steady-state conditions required for comparison with the FEM simulations.

\subsection{Thermal steady state}
The Pt100-measured values and the FEM simulations account for temperature variations on both the left and right sides due to the uneven energy deposition. However, these temperature differences are minimal. Despite the fact that the hottest points are predominantly on the left side of the dump, the temperatures recorded on the right side remain quite similar. Therefore, this analysis primarily focuses on the temperatures observed by the left-side sensors.

\begin{figure}[ht]
    \centering
\includegraphics[width=0.48\textwidth]{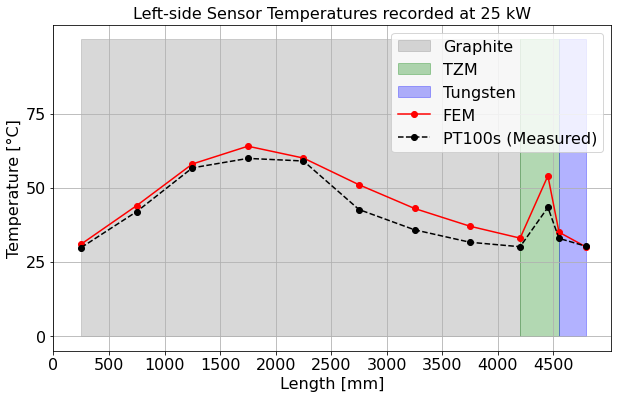}
    \caption{Plots of measured and simulated left-side temperatures in the absorbing blocks at an average thermal power of 25~kW.}
    \label{ss1}
\end{figure}

\begin{figure}[ht]
    \centering
    \includegraphics[width=0.48\textwidth]{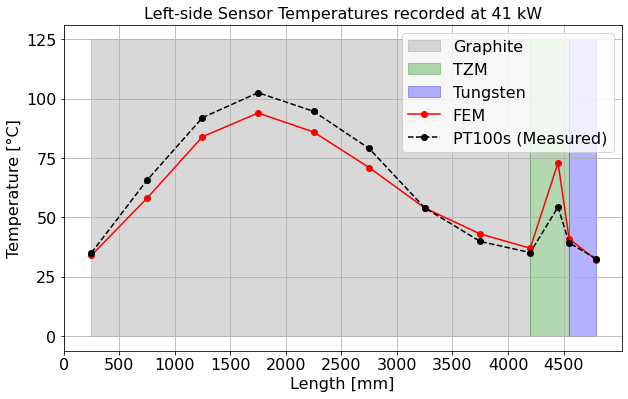}
    \caption{Plots of measured and simulated left-side temperatures in the absorbing blocks at an average thermal power of 41~kW.}
    \label{ss2}
\end{figure}

\begin{figure}[ht]
    \centering
    \includegraphics[width=0.48\textwidth]{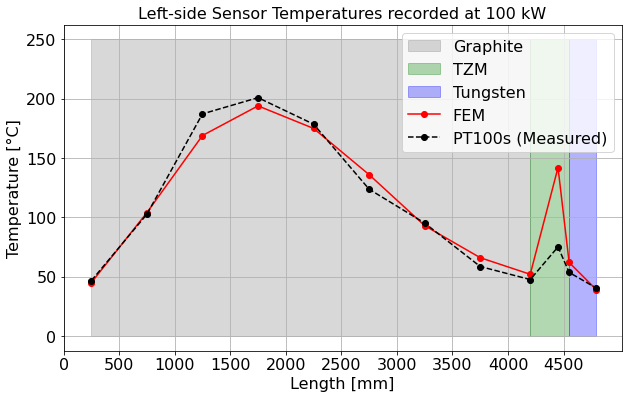}
    \caption{Plots of measured and simulated left-side temperatures in the absorbing blocks at an average thermal power of 100~kW.}
    \label{ss3}
\end{figure}

The plots of measured and simulated temperature in the absorbing blocks in Figs.~\ref{ss1}, \ref{ss2}, and \ref{ss3} correspond to the cases listed in Table~\ref{HRMT}, with average beam thermal powers of 25, 41, and 100~kW, respectively. The values are plotted in relation to the dump's length, such that the segment ``0--4400~mm'' corresponds to the graphite section, ``4400--4600~mm'' to the TZM section, and ``4600--4800~mm'' to the tungsten block. This representation provides valuable insights into the thermal gradients that emerge as the beam traverses the dump, offering a clear understanding of the heat distribution.

\begin{table*}[ht]
    \centering
    \caption{Measured and simulated left-side temperatures under an average power of 25~kW, 41~kW and 100~kW.}
    \begin{tabular}{c c c c c c c} 
        \toprule \toprule
        Sensor location & \multicolumn{2}{c}{25 kW} & \multicolumn{2}{c}{41 kW} & \multicolumn{2}{c}{100 kW} \\
        \cmidrule(lr){2-3} \cmidrule(lr){4-5} \cmidrule(lr){6-7}
        ~ & Pt100 measured ($^{\circ}$C) & FEM ($^{\circ}$C) & Pt100 measured($^{\circ}$C) & FEM ($^{\circ}$C) & Pt100 measured($^{\circ}$C) & FEM ($^{\circ}$C) \\
        \midrule
        CuCr1Zr core & 45.6 & 47.0 & 63.2 & 60.0 & 80.0 & 107.8 \\
        Vacuum chamber 1 & 43.8 & 40.0 & 52.2 & 48.0 & 60.0 & 55.5 \\
        Vacuum chamber 2 & 31.9 & 31.0 & 34.0 & 33.0 & 35.0 & 36.0 \\
        \bottomrule \bottomrule
    \end{tabular}
    \label{PT1001}
\end{table*}

For an average beam thermal power of 25~kW, the FEM model accurately predicts temperatures in the first five graphite blocks, but it tends to overestimate temperatures from the sixth block to the last. Notably, when it comes to the TZM block, the model slightly overestimates temperatures, but they are still within an acceptable range.

In contrast, in the cases of average beam thermal powers of 41 and 100~kW, we observe a reversal in the pattern for the isostatic graphite blocks. In these cases, the temperatures simulated by the FEM model are lower than those measured. The most significant discrepancy occurs in the TZM block, where the model predicts much higher temperatures than those observed. This discrepancy may be attributed to material expansion, which leads to contact with adjacent blocks and subsequent heat dissipation. Another factor could be an overly conservative estimate of the TCC between the TZM blocks and the CuCr1Zr cooling plates.

The measured temperatures of the tungsten block show strong agreement with the numerical values across all three cases, with negligible differences.

To enhance the model, potential improvements could involve reevaluating the specific heat capacity and isotropic thermal conductance of isostatic graphite at high temperatures. Additionally, a reexamination of the TCC between the absorbing blocks and the CuCr1Zr cooling plates may be warranted.

Table~\ref{PT1001} presents temperature data from the CuCr1Zr core sensors and the vacuum chamber in the three different cases. Similar to the absorbing blocks, the temperature variations caused by uneven beam deposition are minimal, with slightly higher temperatures recorded in the left-side sensors.

The temperature measurements were taken at critical locations, including the hottest point of the CuCr1Zr, the upstream region of vacuum chamber~1, and the downstream region of vacuum chamber~2. Notably, there are negligible differences between the Pt100 values and FEM results, indicating a high level of agreement across all average beam thermal powers. However, it is worth mentioning that the FEM model over-predicts the temperature values in the CuCr1Zr core in the 100-kW case.

\subsection{Residual dose rate measurements}

\begin{figure*}[t]
    \centering
\includegraphics[width=0.9\textwidth]{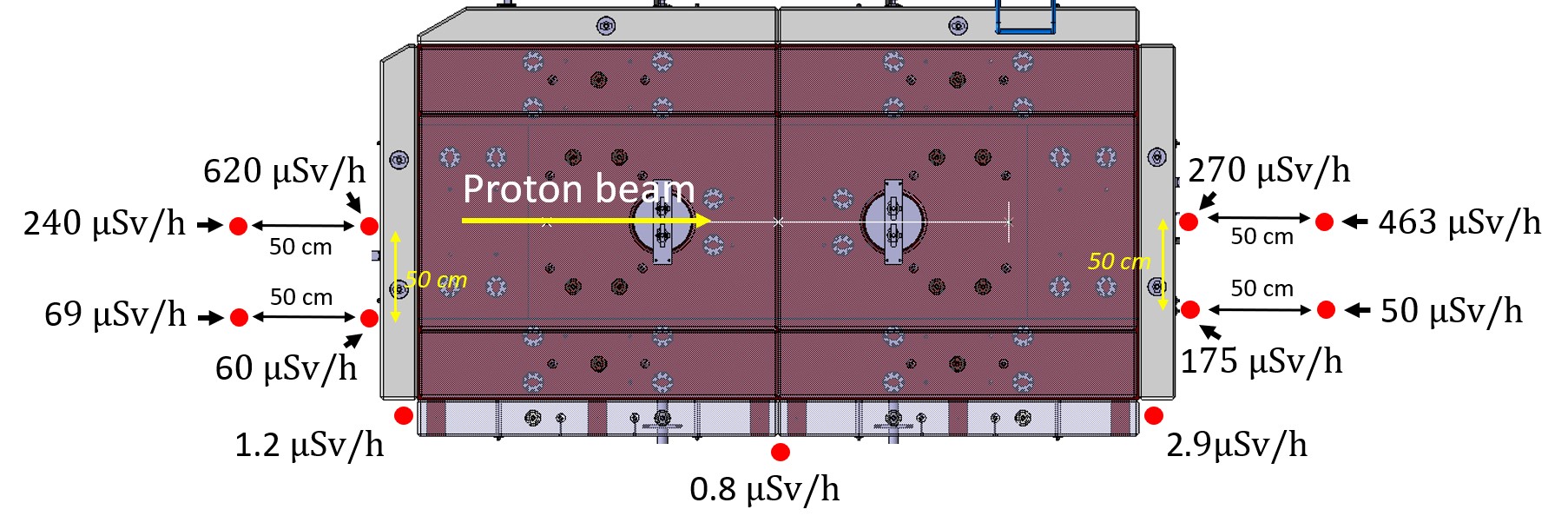}
    \caption{Residual dose rates measured at several points, spanning two years of standard operation followed by a two-day cool-down period. Top view.}
    \label{radiationcont}
\end{figure*}

After a two-day cool-down period following two years of operation, measurements of the residual dose rates in the vicinity of the beam dump were conducted. This was conducted to evaluate the efficacy of both the internal and external shielding and to establish a reference for future comparisons with simulation data. The measurements included three points upstream and three points downstream of the beam dump, as well as one measurement at the center. Two additional measurement points were located 50~cm from the upstream, and further two points were located 50~cm from the downstream region, where the residual dose rate increases due to the masks installed in the area. Figure~\ref{radiationcont} shows the values measured around the external shield.

\section{Conclusions}
To cope with the higher-intensity and more powerful beams that will be operating within the upcoming LIU and HL-LHC era, an innovative and robust new internal beam dump for the CERN SPS, TIDVG\#5, was designed, manufactured, and installed in LSS5 during the CERN LS2, 2019--2020. This new dump will intercept protons accelerated to energy levels from 14 to 450~GeV (the full energy range of the SPS accelerator), and it will be required to withstand a maximum power of $\sim$270~kW delivered by beam pulses, i.e., a fourfold rise compared to the power to which its predecessor was subjected. For this reason, TIDVG\#5 features an upgraded array of absorbing materials providing a higher attenuation factor than TIDVG\#4, HIP-diffusion-bonded CuCr1Zr-316L SS cooling plates for superior thermal-evacuation efficiency, and a seamless and air-cooled vacuum chamber. The change of the dump location from LSS1 to LSS5 enabled a comprehensive redesign of the device, increasing its active length to 5~m and adding massive shielding to allow quick and safe access to the area. The change of position also allowed decoupling of the injection from the dumping system, thus overcoming various difficulties.

The model was validated using three actual cases with average beam thermal powers of 25, 41, and 100~kW deposited in the dump's core. When comparing the operational feedback obtained from the Pt100 sensors and the FEM simulations, the discrepancies found were small, and the temperatures achieved in TIDVG\#5 demonstrated strong agreement with those calculated. This study assessed the reliability of the TIDVG\#5 simulations, and this will also help in the modification or design of future beam dumps.

\section*{Acknowledgments}
The authors would like to acknowledge the LIU-SPS Project at CERN for funding the studies and constructing the system, as well as the multidisciplinary team involved in the project for its significant contribution to the design, production, and installation of this new generation SPS internal beam dump. Specific acknowledgements to M.~Meddahi and B.~Goddard for their support during the entire project. The authors are especially thankful to the following groups at CERN: SY/STI, EN/MME, EN/CV, EN/HE, TE/VSC, BE/CEM, and HSE. The support of the Project Reviewers who provided their time and valuable insights is kindly acknowledged.

\bibliography{TIDVG5_design}

\end{document}